\documentclass[10pt,tightenlines,eqsecnum,floats,aps,amsmath,
amssymb,nofootinbib,prd,notitlepage]{revtex4-1}
\usepackage[utf8x]{inputenc}
\usepackage{bbm}
\usepackage{mathrsfs}
\usepackage{tensor}
\usepackage{array}
\newcommand{\uu}{\mathfrak{u}}
\newcommand{\vv}{\mathfrak{v}}
\newcommand{\ww}{\mathfrak{w}}

\newcommand{\UC}{U${}_{\rm C}$(1)}
\newcommand{\UCUC}{U${}_{\rm C}$(1)$\times$U${}_{\rm C}$(1)}
\newcommand{\UI}{U${}_{\rm I}$(1)}
\newcommand{\UIUI}{U${}_{\rm I}$(1)$\times$U${}_{\rm I}$(1)}
\newcommand{\Tr}[1]{{\rm Tr}\left(#1\right)}

\newcommand{\sJ}[2]{\mathfrak{ J}_{#1}^#2}

\newcommand{\n}{\mathfrak{n}}
\newcommand{\m}{\mathfrak{m}}
\newcommand{\nn}{\mathfrak{q}}
\newcommand{\mm}{\mathfrak{p}}
\newcommand{\nna}[1]{\n_#1}
\newcommand{\mma}[1]{\m_#1}

\newcommand{\F}{\mathcal{F}}

\newcommand{\At}{\mathsf A}
\newcommand{\Atd}{\mathrm A}
\newcommand{\SO}{\mathcal O}
\newcommand{\SOO}{\mathcal S}
\newcommand{\Q}{\mathcal Q}

\newcommand{\G}{\mathcal G}
\newcommand{\ttt}{\mathbf{t}}
\newcommand{\tttt}{\mathfrak{t}}

\newcommand{\lm}{\mathbf{l}}
\newcommand{\lv}{\textbf{\textit{l}}}
\newcommand{\iu}{\mathbbm{i}}
\newcommand{\C}{\mathfrak C}
\newcommand{\DC}[1]{D^C_#1}
\newcommand{\DCC}[1]{\mathfrak{D}_#1}
\newcommand{\Rgt}{\mathfrak{R}}

\newcommand{\kb}{\mathbf k}
\newcommand{\mb}{\mathbf m}
\newcommand{\Pt}{\mathcal{P}^1}
\newcommand{\Pe}{\mathcal{P}^2}
\newcommand{\Ptt}{\mathcal{P}^3}
\newcommand{\Pee}{\mathcal{P}^8}

\newcommand{\Proj}{\texttt{P}}
\newcommand{\proj}{\mathbb{P}}
\newcommand{\x}{u}
\newcommand{\y}{v}
\newcommand{\Mt}{{\mathcal M}^3}
\newcommand{\Lt}{{\mathcal L}^3}
\newcommand{\Kt}{{\mathcal K}^3}
\newcommand{\Me}{{\mathcal M}^8}
\newcommand{\Le}{{\mathcal L}^8}
\newcommand{\Ke}{{\mathcal K}^8}
\newcommand{\omicron}{\mathcal{I}}
\begin{document}
\title{New Faddeev-Niemi type variables for static SU(3) Yang-Mills theory}
\author{Marcin Kisielowski}
\email{Marcin.Kisielowski@fuw.edu.pl}
\affiliation{Instytut Fizyki Teoretycznej, Uniwersytet Warszawski,\\
ul. Ho{\.z}a 69, 00-681 Warszawa (Warsaw), Polska (Poland)}
\affiliation{St. Petersburg Department of Steklov Mathematical Institute,\\
Russian Academy of Sciences, Fontanka 27, St. Petersburg, Russia}

\begin{abstract}
We propose new variables of Faddeev-Niemi type for static SU(3) Yang-Mills theory. These variables reveal a structure of a nonlinear sigma model, whose field variables are two chiral fields taking values in SU(3)/(U(1)xU(1)) and SU(3)/(SU(2)xU(1)). The nonlinear sigma model was introduced by Faddeev and Niemi as a natural extension of the Faddeev $\mathbb{S}^2$ chiral model. Shabanov showed that the energy functional of the extended model is bounded from below by a topological invariant, and therefore may support knot-like excitations and a mass gap.
\end{abstract}
\maketitle
\section{Introduction}
Faddeev and Niemi introduced new variables for the SU(2) Yang-Mills theory \cite{Faddeev_Niemi_SU2}, which reveal a structure of an $\mathbb{S}^2$ nonlinear sigma model \cite{sigma} in the Lagrangian. The nonlinear sigma model is expected to have topologically non-trivial excitations: in the static case, one assumes boundary conditions that compactify the domain to $\mathbb{S}^3$, and therefore the chiral field, being a map from $\mathbb{S}^3$ to $\mathbb{S}^2$, is characterized by the third homotopy group of the sphere $\pi_3(\mathbb{S}^2)$. The corresponding topological charge, given by the Hopf integral, bounds from below the energy functional of the model \cite{VK}. This property suggested, that there are topologically nontrivial solutions of the model. In fact, it has been shown numerically \cite{FNN,Hiet,Hietm,Bat}, that those solutions exist. The application of this model to SU(2) Yang-Mills theory has very interesting consequences. The lower bound supports the belief that the sigma model possesses a mass gap, and 
therefore it is expected, that the Faddeev-Niemi variables support the hypothesis of 
a mass gap in the SU(2) Yang-Mills theory. Moreover, the excitations of the $\mathbb{S}^2$ nonlinear sigma model are knot-like: a pullback of the Kirillov-Konstant symplectic form on $\mathbb{S}^2$ by the map given by the chiral field, can be interpreted as a magnetic field, its lines of force form closed knotted links. These properties suggested, that the particles of Yang-Mills field can be knot-like solitons \cite{FNN,Faddeev_knots}.

It would be particularly interesting to extend the scenario to SU(3) Yang-Mills theory. It is expected, that the gluonic strings are confining quarks. When the quarks are removed, the strings should not disappear. An extension of the Faddeev-Niemi scenario could suggest, that the strings form knotted links. Moreover, the mechanism would support the existence of a mass gap in the SU(3) Yang-Mills theory. One possible extension of the Faddeev-Niemi variables to SU(3) Yang-Mills theory was proposed in \cite{Bolokhov_Faddeev}. It uses the same $\mathbb{S}^2$ sigma model that was used in the SU(2) case. It seems however probable, that also an extension of the sigma model could be important for the SU(3) Yang-Mills theory. A possible extension of the sigma model was suggested in \cite{Faddeev_Niemi_SUN}. In this model, the role of an $\mathbb{S}^2={\rm SU(2)/U(1)}$ chiral field is played by two chiral fields, taking values in SU(3)/(U(1)$\times$U(1)) and SU(3)/(SU(2)$\times$U(1)), i.e. in both orbits of the action of 
the SU(3) group on the Cartan subalgebra of su(3):
$$
n:=-\frac{i}{\sqrt{2}}\, g^\dagger\, \lambda_3\, g, \quad m:=-\frac{i}{\sqrt{2}}\, g^\dagger\, \lambda_8\, g,
$$
where $g:\mathbb{R}^4\to\,$SU(3). The model is defined by an action
\begin{equation}\label{eq:action}
 S=\int d^4\, x\left((\partial_{\mu} n,\partial^{\mu} n)+(\partial_{\mu} m,\partial^{\mu} m)+\frac{1}{e^2}F_{\mu\nu}F^{\mu\nu}+\frac{1}{e^2}G_{\mu\nu}G^{\mu\nu}\right),
\end{equation}
where
\begin{equation}\label{eq:FandG}F_{\mu\nu}=\frac{1}{2}\left( n, [\partial_\mu n, \partial_\nu n]+[\partial_\mu m, \partial_\nu m]\right),\quad G_{\mu\nu}=\frac{1}{2}\left(m,[\partial_\mu n, \partial_\nu n]+[\partial_\mu m, \partial_\nu m]\right)\end{equation} are the Kirillov-Konstant symplectic forms on SU(3)/(U(1)$\times$U(1)) and SU(3)/(SU(2)$\times$U(1)) respectively, $(\cdot,\cdot)$ denotes the SU(3)-invariant scalar product in su(3): $(A,B)=-\Tr{AB}$. The space-time indices $\mu,\nu$ are lowered and raised with Minkowski space-time metric $\eta_{\mu\nu}=\eta^{\mu\nu} = {\rm diag}(-1,1,1,1)$. 
In the static case, one assumes boundary conditions:
$$
n|_\infty=[0,0,1,0,0,0,0,0],\quad m|_\infty=[0,0,0,0,0,0,0,1],
$$
which compactify the domain to $\mathbb{S}^3$. Since \cite{Faddeev_Niemi_SUN}
$$
\pi_3 (\rm{SU(3)/(U(1)}\times\rm{U(1))})=\mathbb{Z},
$$
the model may have knot-like excitations. In fact in \cite{Shabanov} it was shown, that the static energy functional corresponding to the action \eqref{eq:action} is bounded from below by the corresponding topological charge. This supports the belief that the nonlinear sigma model has topologically non-trivial excitations, and a mass gap. Our expectation was, that a certain change of variables may reveal a structure of the extended nonlinear sigma model in the SU(3) Yang-Mills theory in a similar way the Faddeev-Niemi variables reveal the structure of the $\mathbb{S}^2$=SU(2)/U(1) model.
\subsection{The case studied and the structure of the article}
In this article we investigate a static limit of the SU(3) Yang-Mills Lagrangian. By the static limit we mean the case such that
\begin{itemize}
 \item time derivative of the connection $A$ vanishes, i.e. $\partial_0 A_\mu\equiv0$,
 \item time component of the connection $A$ also vanishes, i.e. $A_0\equiv0$.
\end{itemize}
The static Lagrangian can be interpreted as an energy density in the limit of vanishing momenta. The static limit gives an indication on the properties of the ground state of the theory.

The article is structured as follows. In section \ref{sc:SU2_variables} we present an interpretation of the Faddeev-Niemi decomposition, which will justify our choice of variables for the SU(3) theory. In section \ref{sc:variables} we introduce our variables for the SU(3) Yang-Mills theory and discuss the relation with the chiral model \eqref{eq:action}. Thanks to structural similarity between our decomposition and Faddeev-Niemi decomposition for SU(2) Yang-Mills theory many calculations in the SU(3) case are analogous to calculations in the SU(2) case. We summarize the main properties of the models in section \ref{sc:summary}, and then in section \ref{sc:discussion} we discuss the results and give an outlook on further research.
\section{The static SU(2) Faddeev-Niemi variables}\label{sc:SU2_variables}
In this section we present an interpretation of the static SU(2) Faddeev-Niemi variables, which justifies the decomposition we introduce in the next section.
\subsection{The decomposition of the connection}\label{sc:decomposition}
In the original paper Faddeev and Niemi decomposed the SU(2) connection $\At_\mu=\At^p_\mu \sigma^p, p\in\{1,2,3\}$ in the Cartan-Weyl basis of the su(2) Lie algebra:
$$
\At_\mu=\Atd_\mu \sigma^3 + X^+_\mu \sigma^-+X^-_\mu \sigma^+,
$$
where $\sigma^{\pm}=\frac{1}{2}\left(\sigma^1 \pm i \sigma^2 \right)$, and $\sigma^1,\ \sigma^2,\ \sigma^3$ are the Pauli matrices.

As we consider only the static case, we discuss here only the static counterpart of the Faddeev-Niemi decomposition. 
We interpret $X^+_j(x)$ as components of a co-vector:
$$
X^+:=\left[X^+_1,X^+_2,X^+_3 \right].
$$
We consider an action of an SO(3) group on this co-vector:
$$
\SO \cdot X^+ := X^+ \SO^T,\quad \SO\in {\rm SO(3)}.
$$
 The idea of the Faddeev-Niemi decomposition is based on the following observation: if the vectors ${\rm Re}\left(X^+ \right)$ and ${\rm Im}\left(X^+ \right)$ are not co-linear, then there exists an SO(3) group element that sets the third co-ordinate to zero:
\begin{equation}\label{eq:SOX}
\SO \cdot X^+ = \left[ \Phi_1 , \Phi_2 , 0 \right].
\end{equation}
The construction is justified by the fact, that $\At^1_i$ and $\At^2_i$ can be expressed in a orthonormal basis $e^1_i$, $e^2_i$ of the plane spanned by $\At^1_i$ and $\At^2_i$:
\begin{equation}\label{eq:XPe}
X^+=\At^1+i\,\At^2=\Phi_1 e^1 + \Phi_2 e^2.
\end{equation}
This means, that the matrix $\SO$ is the SO(3) matrix, whose first two rows are $e^1$ and $e^2$, respectively, and the third row is $e^1\times e^2$, where $\times$ denotes the vector product. 

We denote:
$$
\Phi:= \left[ \Phi_1 , \Phi_2 , 0 \right],
$$
 and decompose $\Phi$ into a normed vector 
$$\Q=[\phi_1,\phi_2,0]:=\frac{1}{\sqrt{|\Phi_1|^2+|\Phi_2|^2}}\left[ \Phi_1 , \Phi_2 , 0 \right]$$
and a scalar field $\rho=\sqrt{|\Phi_1|^2+|\Phi_2|^2}$. The decomposition of $X^+$ takes now the following form:
\begin{equation}\label{eq:decomposition_SU2}X^+ = \Q\, \rho\, \SO.\end{equation}
Importantly, note that $\Q$ uniquely defines an SU(2) matrix (see also \cite{Faddeev_WS}):
$$
Q=\left(\begin{array}{cc} \phi_1 & -\overline{\phi}_2\\ \phi_2& \overline{\phi}_1\end{array} \right).
$$
\subsection{Gauge transformations}
\subsubsection{External gauge transformations}
There is a U(1) symmetry of the action called external gauge symmetry and denoted by \UC. The symmetry comes from the fact, that the Maximal Abelian Gauge fixing is not fixing the gauge completely. The remaining gauge transformations act on $Q$ by right translations. The corresponding covariant derivative is:
$$
D_i Q := \partial_i Q + \iu \Atd_i \, Q\, \sigma_3, \quad D_i Q^\dagger := \partial_i Q^\dagger - \iu \Atd_i\, \sigma_3\, Q^\dagger.
$$
\subsubsection{Internal gauge transformations}
There is also a U(1) symmetry of the action arising from a U(1) ambiguity in the decomposition \eqref{eq:decomposition_SU2}. The corresponding gauge transformations are called internal gauge transformations \cite{Faddeev_Niemi_SU2} and denoted by \UI. The ambiguity comes from the fact, that the SO(3) group element $\SO$ and the vector $\Phi$ \eqref{eq:SOX} are not defined uniquely. For 
$$
\G :=\left(\begin{array}{ccc} \cos \lambda & -\sin\lambda&0\\ \sin\lambda & \cos\lambda &0\\0&0&1\end{array} \right)
$$
both matrices $\SO$ and $\G \SO$ set the third component of $X^+$ to zero. Explicitly, the symmetry is:
$$
\SO \mapsto \G^{-1} \SO, \quad \Phi \mapsto \Phi \G.
$$
In other words, the \UI\ ambiguity is the ambiguity in the choice of the orthonormal basis $e^1_i$, $e^2_i$ \eqref{eq:XPe}. The \UI\ gauge transformations act on $Q$ and on $\SO$ by right translations. In \cite{Faddeev_Niemi_SU2} a gauge potential $C_i$ corresponding to these transformations was constructed. In our notation it takes the following form:
\begin{equation}\label{eq:Ci}
C_i:=\frac{1}{2} \Tr{\epsilon_3 \SO \partial_i \SO^T},
\end{equation}
where  $\epsilon_1,\ \epsilon_2,\ \epsilon_3$ denote the matrices of the spin 1 representation of the SU(2) Lie algebra:
\begin{equation}\label{eq:epsilon}
(\epsilon_k)_{ij}:= \epsilon_{kij}, \quad k,i,j\in\{1,2,3\},
\end{equation}
$\epsilon_{ijk}$ is the Levi-Civita symbol. 
The derivative covariant with respect to \UI\ and \UC\ gauge transformations is:
\begin{equation}\label{eq:DC}
\DC{i} Q = D_i Q - \iu C_i \sigma_2 Q,\quad \DC{i} Q^\dagger = D_i Q^\dagger + \iu C_i  Q^\dagger\sigma_2 .
\end{equation}
\subsection{Chiral fields and supercurrent}
The SU(2) Yang-Mills action is further interpreted in \cite{Faddeev_Niemi_SU2} in terms of chiral fields and supercurrents. We define those fields in the following way.
\subsubsection{Chiral fields}
The first chiral field is associated with the $Q$ variable:
\begin{equation*}\label{eq:sigmaQ}
\tttt:=\frac{1}{\sqrt{2}}Q \sigma_3 Q^\dagger.
\end{equation*}
We denote its coordinates by $\tttt_i,\ i\in{1,2,3}$:
$$
\tttt_i:=\frac{1}{\sqrt{2}}\Tr{\sigma_i \tttt}.
$$
We will identify $\tttt$ with a unit vector field in $\mathbb{R}^3$ whose components are $\tttt_i$. We will use a same notation for the vector field, and the corresponding su(2) valued field. It should cause no confusion, because the formulas will involve either matrix operations (e.g. a trace Tr or a commutator $[\cdot,\cdot]$) or vector operations (e.g. a scalar product $\cdot$ or a vector product $\times$). 

Second chiral field comes from the matrix $\SO(x)$. It is defined in the following way:
\begin{equation}\label{eq:l}
\lm= \frac{1}{\sqrt{2}} \SO^T \epsilon_3 \SO.
\end{equation}
Coordinates of this vector read:
$$
l_i:=-\frac{1}{\sqrt{2}} \Tr{\epsilon_i \lm}.
$$
We denote by $\lv$ a normalised vector field in $\mathbb{R}^3$ whose components are $l_i$. It is easy to check, that this vector field coincides with the field $\lv$ in \cite{Faddeev_Niemi_SU2}. However the field $\tttt$ is different from $\ttt$ defined in \cite{Faddeev_Niemi_SU2}. We discuss this issue in subsection \ref{sc:original_FN}.

\subsubsection{Supercurrent}
We define the supercurrent by the following equation:
\begin{equation}\label{eq:supercurrentSU2}
J_i=\frac{\iu}{2} \Tr{\sigma_3 Q^\dagger \DC{i} Q}.
\end{equation}
It is equal to the \UI$\times$\UC\ invariant supercurrent from \cite{Faddeev_Niemi_SU2}.
\subsection{The static Lagrangian}
Further, Faddeev and Niemi re-express the Lagrangian of the SU(2) Yang-Mills theory in the new variables (chiral fields and supercurrents). They partially fix the gauge using Maximal Abelian Gauge fixing condition. The static SU(2) Yang-Mills Lagrangian together with the gauge fixing terms and with ghost terms omitted is \cite{Faddeev_Niemi_SU2}:
\begin{equation}\label{eq:YM_action}
L_{SU(2)} = \frac{1}{4}( F_{ij} + 2 P_{ij})^2  + \  \frac{1}{2} |\mathrm D_{i} X^+_j|^2 
-\frac{3}{4} P_{ij}^2,
\end{equation}
where $$D_i X^\pm_j = (\partial_i \pm i \Atd_i ) X^\pm_j,\quad F_{ij}=\partial_i \Atd_j - \partial_j \Atd_i, \quad P_{ij}=\frac{\iu}{2}\left( X^+_i X^-_j - X^+_j X^-_i\right).$$

\subsection{The term $\frac{1}{2}|D_i X^+_j|^2$}\label{sc:quadratic_term}
The static Lagrangian can be naturally divided into two parts. The first part is $\frac{1}{2}|D_i X^+_j|^2$. It contains the terms of standard sigma model Hamiltonian for both chiral fields.
\subsubsection{Applying the decomposition}
First, we rewrite the term $\frac{1}{2}|D_i X^+_j|^2$ using the decomposition $X^+=\Q \rho \SO$ introduced in subsection \ref{sc:decomposition}:
$$
\frac{1}{2}|D_i X^+_j|^2=\frac{1}{2}\left(D_i\, \Q\, \rho\, \SO + \Q\, \partial_i\, \rho\, \SO + \Q\, \rho \,\partial_i\SO\right)\cdot\left(\SO^T \rho\, D_i \Q^\dagger + \SO^T \partial_i \rho\, \Q^\dagger + \partial_i\SO^T \rho\, \Q^\dagger \right),
$$
where $\cdot$ denotes a natural action of a co-vector on a vector. It is now an easy calculation to show, that:
\begin{eqnarray}
\frac{1}{2}|D_i X^+_j|^2=\frac{\rho^2}{2} D_i\, \Q\cdot D_i \Q^\dagger+ \frac{1}{2} (\partial_i \rho)^2+\frac{\rho^2}{2}\Tr{(\Q^\dagger\otimes \Q) \partial_i \SO \partial_i \SO^T}+&\nonumber\\+\frac{\rho^2}{2}\Tr{(\Q^\dagger \otimes D_i \Q)\SO \partial_i \SO^T}&-\frac{\rho^2}{2}\Tr{(D_i\Q^\dagger \otimes \Q)\SO \partial_i \SO^T},\label{eq:DXUC}
\end{eqnarray}
where
$$
\Q^\dagger\otimes \Q=\left(\begin{array}{ccc}
                            |\phi_1|^2& \overline{\phi}_1\phi_2&0\\
			    \overline{\phi}_2 \phi_1& |\phi_2|^2&0\\
			    0&0&0
                           \end{array}
 \right), \quad \Q^\dagger\otimes D_i \Q=\left(\begin{array}{ccc}
                            \overline{\phi}_1 D_i \phi_1 & \overline{\phi}_1D_i\phi_2&0\\
			    \overline{\phi}_2 D_i \phi_1& \overline{\phi}_2 D_i \phi_2&0\\
			    0&0&0
                           \end{array}
 \right),\quad  D_i\Q^\dagger\otimes \Q=\left(\begin{array}{ccc}
                            D_i\overline{\phi}_1  \phi_1 & D_i\overline{\phi}_1\phi_2&0\\
			    D_i\overline{\phi}_2 \phi_1& D_i\overline{\phi}_2 \phi_2&0\\
			    0&0&0
                           \end{array}
 \right),
$$
$D_i \phi_1 = \partial_i \phi_1 + \iu \Atd_i \phi_1,\ D_i \phi_2 = \partial_i \phi_2 + \iu \Atd_i \phi_2$. Thanks to the properties of covariant derivative, each term in the expression \eqref{eq:DXUC} is \UC\ gauge invariant. However, not every term is \UI\ gauge invariant, and only the whole expression is. In next subsection we rewrite it as a sum of terms that are \UI\ gauge invariant; we call this expression an explicitly gauge invariant form.
\subsubsection{Explicitly gauge invariant form}\label{sc:eginv_form}
Note, that $\SO \partial_i \SO^T$ takes values in the Lie algebra of SO(3) which is spanned by $\epsilon_1,\ \epsilon_2,\ \epsilon_3$. Since the third row and the third column of $\Q^\dagger\otimes D_i \Q$ vanish, only the component corresponding to $\epsilon_3$ is nonzero.
 Therefore
$$
\rho^2\Tr{(\Q^\dagger \otimes D_i \Q)\SO \partial_i \SO^T}=-\frac{1}{2}\rho^2\Tr{\epsilon_3(\Q^\dagger \otimes D_i \Q)}\Tr{\epsilon_3\SO \partial_i \SO^T}=-C_i\rho^2 D_i \Q \epsilon_3 \Q^\dagger.
$$
Similarly,
$$
\rho^2\Tr{(D_i \Q^\dagger \otimes \Q)\SO \partial_i \SO^T}=-C_i\rho^2  \Q \epsilon_3 D_i\Q^\dagger.
$$
As a result:
\begin{equation}\label{eq:DQDQCC}
\frac{\rho^2}{2} D_i\, \Q\cdot D_i \Q^\dagger+\frac{\rho^2}{2}\Tr{(\Q^\dagger \otimes D_i \Q)\SO \partial_i \SO^T}-\frac{\rho^2}{2}\Tr{(D_i\Q^\dagger \otimes \Q)\SO \partial_i \SO^T}=\frac{\rho^2}{2} \DC{i}\, \Q\cdot \DC{i} \Q^\dagger - \frac{\rho^2}{2}C_i^2,
\end{equation}
where the covariant derivatives $\DC{i}\Q$ and $\DC{i}\Q^\dagger$ are the covariant derivatives compatible with \eqref{eq:DC}:
$$
\DC{i}\Q=\left[\DC{i}\phi_1,\DC{i}\phi_2,0\right],\ \DC{i}\Q^\dagger=(\DC{i}\Q)^\dagger,\ \DC{i}\phi_1=\partial_i \phi_1 + \iu \Atd_i \phi_1 - C_i \phi_2,\ \DC{i}\phi_2=\partial_i \phi_2 + \iu \Atd_i \phi_2 + C_i \phi_1.
$$
Using the fact, that $\SO \SO^T=\SO^T \SO =1$, one can show, that
$$
\frac{\rho^2}{2}\Tr{(\Q^\dagger\otimes \Q) \partial_i \SO \partial_i \SO^T}=-\frac{\rho^2}{2}\Tr{(\Q^\dagger\otimes \Q) \SO \partial_i \SO^T \SO \partial_i \SO^T}.
$$
Using the fact, that 
\begin{equation}\label{eq:ee}
(\epsilon_i \epsilon_j)_{mn}= \epsilon_{imm'} \epsilon_{jm'n}=-\delta_{ij}\delta_{mn}+\delta_{in}\delta_{mj},
\end{equation}
and the fact that the third column and the third row of the matrix $\Q^\dagger\otimes \Q$ are equal zero, one can show, that:
\begin{equation}\label{eq:QQ}
\Tr{(\Q^\dagger\otimes \Q) \SO \partial_i \SO^T \SO \partial_i \SO^T}=\Tr{(\Q^\dagger\otimes \Q) \proj( \SO \partial_i \SO^T) \proj( \SO \partial_i \SO^T)}-C_i^2,
\end{equation}
where $\proj:so(3)\to so(3)$ is an orthogonal projector onto the subspace spanned by $\epsilon_1$ and $\epsilon_2$. From equations \eqref{eq:DQDQCC} and \eqref{eq:QQ} follows now, that:
\begin{equation}\label{eq:DXDX}
\frac{1}{2}|D_i X^+_j|^2=\frac{\rho^2}{2} \DC{i}\, \Q\cdot \DC{i} \Q^\dagger+ \frac{1}{2} (\partial_i \rho)^2-\frac{\rho^2}{2}\Tr{(\Q^\dagger\otimes \Q) \proj( \SO \partial_i \SO^T) \proj( \SO \partial_i \SO^T)}.
\end{equation}
 Importantly, note that $\proj (\SO \partial_i \SO^T)$ is \UI\ gauge covariant; under \UI\ gauge transformations it transforms in the following way:
$$
\proj (\SO \partial_i \SO^T)\mapsto \G^{-1} \proj (\SO \partial_i \SO^T) \G.
$$
Knowing this, it is straightforward to check, that each term in the expression \eqref{eq:DXDX} is \UC$\times$\UI\ gauge invariant.
\subsubsection{The term $\frac{\rho^2}{2} \DC{i}\, \Q\cdot \DC{i} \Q^\dagger$}
The term $ \frac{\rho^2}{2} \DC{i}\, \Q\cdot \DC{i} \Q^\dagger $ can be interpreted in terms of the chiral field $\tttt$ and the supercurrent $J_i$. First, notice that
$$
\frac{\rho^2}{2} \DC{i}\, \Q\cdot \DC{i} \Q^\dagger = \frac{\rho^2}{4} \Tr{\DC{i}Q^\dagger \DC{i}Q}=-\frac{\rho^2}{4} \Tr{Q\DC{i}Q^\dagger Q\DC{i}Q^\dagger}.
$$

Using the fact, that $\frac{1}{2}[\tttt,[\tttt,\cdot]]$ is an orthogonal projection onto the space orthogonal to $\tttt$, and the fact that the orthogonal projection of $Q\DC{i} Q^\dagger$ onto the space spanned by $\tttt$ is $\iu \sqrt{2} J_i$, it can be shown, that (compare with e.g. \cite{Shabanov_UdU,Faddeev_Niemi_SUN}):
\begin{equation}\label{eq:QdQtDt}
Q\DC{i} Q^\dagger=\frac{1}{2}[\tttt,\DC{i} \tttt]+ \iu \sqrt{2} J_i \tttt,
\end{equation}
where $\DC{i}\tttt = \partial_i \tttt - \iu C_i [\sigma_2,\tttt] $. As a result:
\begin{equation}\label{eq:DCQDCQ}
\frac{\rho^2}{2} \DC{i}\, \Q\cdot \DC{i} \Q^\dagger =\frac{\rho^2}{8} (\DC{i} \tttt)^2 + \frac{\rho^2}{2} (J_i)^2.
\end{equation}
When $\rho$ is interpreted as a condensate, the term $\frac{\rho^2}{8} (\DC{i} \tttt)^2$ becomes a standard sigma model Hamiltonian, and the term $\frac{\rho^2}{2} (J_i)^2$ becomes a mass term for the supercurrent $J_i$.
\subsubsection{The term $-\frac{\rho^2}{2}\Tr{(\Q^\dagger\otimes \Q) \proj(\SO \partial_i \SO^T)\proj( \SO \partial_i \SO^T)}$}\label{sc:QQDODO}
This term gives rise to standard sigma model Hamiltonian term for the chiral field $\lv$. One may show, that:
$$
\Tr{(\Q^\dagger\otimes \Q) \proj(\SO \partial_i \SO^T)\proj( \SO \partial_i \SO^T)}=\frac{1}{4}\Tr{\proj(\SO \partial_i \SO^T)\proj( \SO \partial_i \SO^T)}+\frac{1}{2}\left(\Rgt_i^1 \Rgt_i^1-\Rgt_i^2\Rgt_i^2 \right)\tttt_3+ \Rgt_i^1 \Rgt_i^2 \tttt_1,
$$
where $\Rgt_i = \SO \partial_i \SO^T$ is a pullback of the right-invariant form on SO(3) and $\Rgt_i^p$ are its components in the basis $\epsilon_1,\,\epsilon_2,\,\epsilon_3$:
$$
\Rgt_i =\Rgt_i^p \epsilon_p,\ \Rgt_i^p =-\frac{1}{2}\Tr{\epsilon_p \SO \partial_i \SO^T},\ p\in\{1,2,3\}.
$$
Note, that:
$$
\proj(\SO \partial_i \SO^T)=-[\epsilon_3,[\epsilon_3,\SO \partial_i \SO^T]].
$$
Therefore
$$
\SO^T \proj(\SO \partial_i \SO^T) \SO =2[\lm,[\lm,\SO^T \partial_i \SO]].
$$
It is easy to show, that:
$$
\partial_i \lm = [\lm, \SO^T \partial_i \SO].
$$
As a result:
$$
\SO^T \proj(\SO \partial_i \SO^T) \SO =2[\lm,\partial_i\lm].
$$
We obtain:
\begin{equation}\label{eq:dldl}
\Tr{ \proj(\SO \partial_i \SO^T)\proj( \SO \partial_i \SO^T)}=4\,\Tr{[\lm,\partial_i\lm][\lm,\partial_i\lm]}=2 \Tr{\partial_i \lm \partial_i \lm}=-2 \partial_i l_j \partial_i l_j=-2 (\partial_i \lv )^2,
\end{equation}
and
\begin{eqnarray*}
\Rgt_i^1=-\frac{1}{2}\Tr{\epsilon_1 \SO \partial_i \SO^T}=-\frac{1}{2}\Tr{\epsilon_1 \proj(\SO \partial_i \SO^T)}=-\Tr{\SO^T\epsilon_1\SO[\lm,\partial_i\lm]},\\
\Rgt_i^2=-\frac{1}{2}\Tr{\epsilon_2 \SO \partial_i \SO^T}=-\frac{1}{2}\Tr{\epsilon_2 \proj(\SO \partial_i \SO^T)}=-\Tr{\SO^T\epsilon_2\SO[\lm,\partial_i\lm]}.
\end{eqnarray*}
We introduce two auxiliary unit vectors:
\begin{equation}\label{eq:km}
\kb=\frac{1}{\sqrt{2}}\SO^T \epsilon_1 \SO ,\quad \mb=\frac{1}{\sqrt{2}}\SO^T \epsilon_2 \SO.
\end{equation}
This time $\kb$ and $\mb$ are analogous but in general not equal to $\textbf{\textit{k}}$ and $\textbf{\textit{m}}$ from \cite{Faddeev_Niemi_SU2} (however, they lie in the same plane, and therefore at each point $\kb$ and $\mb$ are related to $\textbf{\textit{k}}$ and $\textbf{\textit{m}}$ by an SO(2) rotation). This will be the reason for the form of the term $\frac{1}{2}\left(\Rgt_i^1 \Rgt_i^1-\Rgt_i^2\Rgt_i^2 \right)\tttt_3+ \Rgt_i^1 \Rgt_i^2 \tttt_1$ to be different from the corresponding one in \cite{Faddeev_Niemi_SU2}. 

Note, that:
$$
[\kb,\lm]= \frac{1}{\sqrt{2}} \mb,\quad [\lm,\mb]=\frac{1}{\sqrt{2}}\kb, \quad [\mb,\kb]=\frac{1}{\sqrt{2}}\lm.
$$
Therefore:
$$
\Rgt_i^1=-\sqrt{2}\Tr{\kb[\lm,\partial_i\lm]}=-\Tr{\mb\partial_i\lm},\quad \Rgt_i^2=-\sqrt{2}\Tr{\mb[\lm,\partial_i\lm]}=\Tr{\kb\partial_i\lm}.
$$
Finally:
$$
-\frac{\rho^2}{2}\Tr{(\Q^\dagger\otimes \Q) \proj(\SO \partial_i \SO^T)\proj( \SO \partial_i \SO^T)}=\frac{\rho^2}{4} (\partial_i \lv)^2+\frac{\rho^2}{4}\left((\Tr{\kb\partial_i \lm})^2-(\Tr{\mb\partial_i \lm})^2 \right)\tttt_3+\frac{\rho^2}{2} \Tr{\kb\partial_i\lm}\Tr{\mb\partial_i\lm} \tttt_1.
$$
The term gives rise to the standard sigma model Hamiltonian $(\partial_i \lv)^2$ of the chiral field $\lv$.
\subsubsection{Final expression for the term $\frac{1}{2}|D_i X^+_j|^2$}
Taking into account considerations from the previous subsections, one can write the investigated term in the following form:
$$
\frac{1}{2}|D_i X^+_j|^2=\frac{1}{2} (\partial_i \rho)^2+\frac{\rho^2}{8} (\DC{i} \tttt)^2 + \frac{\rho^2}{2} (J_i)^2+\frac{\rho^2}{4} (\partial_i \lv)^2+\frac{\rho^2}{4}\left((\Tr{\kb\partial_i \lm})^2-(\Tr{\mb\partial_i \lm})^2 \right)\tttt_3+\frac{\rho^2}{2} \Tr{\kb\partial_i\lm}\Tr{\mb\partial_i\lm} \tttt_1.
$$
The term $\frac{1}{2}|D_i X^+_j|^2$ gives rise to the two standard sigma model Hamiltonians $(\DC{i} \tttt)^2$, $(\partial_i \lv)^2$ and terms $\frac{\rho^2}{2} (J_i)^2$, which make the supercurrent $J_i$ massive.
\subsection{The tensor $P_{ij}$}\label{sc:P}
 Let us recall, that
$$
P_{ij}=\frac{\iu}{2}\left(X^+_i X^-_j - X^+_j X^-_i \right).
$$
The tensor $P_{ij}$ describes an embedding of the real oriented Grassmanian $\widetilde{\rm G}$(3,2)=SO(3)/SO(2) into the three-dimensional space of two-forms in $\mathbb{R}^3$, i.e. in $\bigwedge^2 \mathbb{R}^3$.
It will be convenient for us to consider the tensor
$$
P_k:=\epsilon_{ijk} P_{ij}=\iu \epsilon_{ijk} X^+_i X^-_j.
$$
instead of the tensor $P_{ij}$. Obviously, the relation can be inverted:
$$
P_{ij}=\frac{1}{2}\epsilon_{ijk} P_k.
$$
Using the matrices $\epsilon_i$ introduced in \eqref{eq:epsilon} we write:
$$
P_k=\iu X^+ \epsilon_k (X^+)^\dagger.
$$
Applying the decomposition \eqref{eq:decomposition_SU2}, we obtain:
$$
P_k=\iu  \Q\, \rho\, \SO\, \epsilon_k\, \SO^T \rho\, \Q^\dagger = \iu\rho^2 \Tr{(\Q^\dagger \otimes \Q) \SO\, \epsilon_k\, \SO^T}.
$$
Since $\SO\, \epsilon_k\, \SO^T$ has values in the so(3) Lie algebra, only the components of $\Q^\dagger \otimes \Q$ corresponding to $\epsilon_1,\epsilon_2,\epsilon_3$ contribute. Since the third row and third column of $\Q^\dagger \otimes \Q$ are zero, only the components corresponding to $\epsilon_3$ contribute. Therefore
$$
P_k= \rho^2 \tttt_2 \lm_k.
$$

As a result
$$
P_{ij}=\frac{1}{2}\epsilon_{ijk} \rho^2 \tttt_2 l_k.
$$
This is an explicit form of the embedding: $l_k$ are coordinates of a normalized vector field $\lv$ in $\mathbb{R}^3$, $\lv(x)\in \mathbb{S}^2=SO(3)/SO(2)=\widetilde{\rm G}(3,2)$.
In the expression for the static SU(2) Yang-Mills Lagrangian also $(P_{ij})^2$ appears. Obviously, it does not depend on the field $\lv$:
$$
P_{ij} P_{ij} = \frac{1}{2} P_k P_k = \frac{1}{2} \rho^4 \tttt_2^2.
$$
\subsection{The Maxwell tensor $F_{ij}$}
Further, Faddeev and Niemi replace the \UC\ gauge potential $\Atd_i$ with the \UC$\times$\UI\ gauge invariant supercurrent $J_i$.  From the definition \eqref{eq:supercurrentSU2} of the supercurrent $J_i$ follows, that
$$
J_i=\frac{\iu}{2}\Tr{\sigma_3 Q^\dagger \partial_i Q}-\Atd_i + \tttt_2 C_i.
$$
Treating now $J_i$ as independent variable, we obtain:
$$
\Atd_i=\frac{\iu}{2}\Tr{\sigma_3 Q^\dagger \partial_i Q}-J_i + \tttt_2 C_i.
$$

The Maxwell tensor is:
$$
F_{ij}=\frac{\iu}{2}\partial_{[i}\Tr{\sigma_3 Q^\dagger \partial_{j]} Q}-\partial_{[i} J_{j]} + \partial_{[i}\left( \tttt_2 C_{j]}\right).
$$
The term $\partial_{[i}\Tr{\sigma_3 Q^\dagger \partial_{j]} Q}$ is a pullback of the Kirillov-Konstant symplectic form on $\mathbb{S}^2$. Expressed in terms of the $\mathbb{S}^2$ valued field $\tttt$ it takes the following form:
$$
\partial_{[i}\Tr{\sigma_3 Q^\dagger \partial_{j]} Q}=\frac{\sqrt{2}}{2}\Tr{\tttt\,[\partial_i \tttt, \partial_j \tttt]}.
$$
Therefore:
$$
F_{ij}=\frac{\iu \sqrt{2}}{4}\Tr{\tttt\,[\partial_i \tttt, \partial_j \tttt]}-\partial_{[i} J_{j]} + \partial_{[i}\left( \tttt_2 C_{j]}\right).
$$
Further, it may be shown, that:
$$
F_{ij}=\frac{\iu \sqrt{2}}{4}\Tr{\tttt\,[\DC{i} \tttt, \DC{j} \tttt]}+\tttt_2 \partial_{[i} C_{j]}-\partial_{[i} J_{j]}.
$$
Note, that each term in this formula is \UI\ gauge invariant. The imaginary unit in this expression may be surprising at first sight. However, let us remind that we use a matrix notation in this formula. In vector notation the imaginary unit disappears:
$$
F_{ij}=-\frac{1}{2}\tttt \cdot (\DC{i}\tttt \times \DC{j} \tttt)+\tttt_2 \partial_{[i} C_{j]}-\partial_{[i} J_{j]},
$$
where $\DC{i}\tttt_j=\partial_i \tttt_j-2 C_i \epsilon_{2jk}\tttt_k$. The term $\partial_{[i} C_{j]}$ can be also interpreted using the $\lm$ field (equivalently the $\lv$ field):
$$
\partial_{[i} C_{j]}=\sqrt{2} \Tr{\lm [\partial_i \lm,\partial_j \lm]}=-\lv\cdot (\partial_i \lv \times \partial_j\lv),
$$
where $-\lv\cdot (\partial_i\lv \times \partial_j\lv)$ is again a pullback of the Kirillov-Konstant symplectic form on $\mathbb{S}^2$.
As a result the Maxwell tensor has the following interpretation:
$$
F_{ij}=-\frac{1}{2}\tttt \cdot (\DC{i}\tttt \times \DC{j} \tttt)-\lv\cdot (\partial_i \lv \times \partial_j\lv)\,\tttt_2  -\partial_{[i} J_{j]}.
$$
It provides the Maxwell terms $(\tttt \cdot (\DC{i}\tttt \times \DC{j} \tttt))^2$ and $(\lv\cdot (\partial_i \lv \times \partial_j\lv))^2$ for the chiral fields $\tttt$ and $\lv$. Together with the standard sigma model Hamiltonian terms, they describe how the $\mathbb{S}^2$ sigma model \cite{sigma} appears in the Lagrangian.
\subsection{The static Lagrangian in Faddeev-Niemi variables}
We can write now the static Lagrangian \eqref{eq:YM_action} in the following form
\begin{eqnarray}
&L_{SU(2)} = \frac{1}{4}\left(\frac{1}{2}\tttt \cdot (\DC{i}\tttt \times \DC{j} \tttt)+(\lv\cdot (\partial_i \lv \times \partial_j\lv)-\rho^2 \epsilon_{ijk}l_k)\,\tttt_2  +\partial_{[i} J_{j]} \right)^2 + \frac{1}{2} (\partial_i \rho)^2 -\frac{3}{8}\rho^4 \tttt_2^2+ \nonumber\\&+\frac{\rho^2}{8} (\DC{i} \tttt)^2 + \frac{\rho^2}{2} (J_i)^2+\frac{\rho^2}{4} (\partial_i \lv)^2+\frac{\rho^2}{4}\left((\Tr{\kb\partial_i \lm})^2-(\Tr{\mb\partial_i \lm})^2 \right)\tttt_3+\frac{\rho^2}{2} \Tr{\kb\partial_i\lm}\Tr{\mb\partial_i\lm} \tttt_1.\label{eq:sigma}\end{eqnarray}
This form reveals nonlinear sigma model terms in the static SU(2) Yang-Mills Lagrangian. The chiral fields are $\tttt$ and $\lv$, and both of them may support knot-like excitations. Note that, there is a substantial structural duality between the fields $\tttt$ and $\lv$. The field $\tttt$ is interpreted as ``electric`` and the field $\lv$ as ''magnetic`` order parameter \cite{Faddeev_Niemi_SU2}.

The expression \eqref{eq:sigma} should be compared with formulas (81),(82),(83) and (84) from \cite{Faddeev_Niemi_SU2}. We do not recover exactly the original expression, because there are some minor differences between the variables we use and the original Faddeev-Niemi variables. As we pointed out the main difference is in the expression for the term $\frac{\rho^2}{4}\left((\Tr{\kb\partial_i \lm})^2-(\Tr{\mb\partial_i \lm})^2 \right)\tttt_3+\frac{\rho^2}{2} \Tr{\kb\partial_i\lm}\Tr{\mb\partial_i\lm} \tttt_1$ (compare with equation (84) in \cite{Faddeev_Niemi_SU2}), which is caused by the fact that the fields $\kb$ and $\mb$ we introduced are not in general equal to $\textbf{\textit{k}}$ and $\textbf{\textit{m}}$ from \cite{Faddeev_Niemi_SU2} (at each point they lie in the same plane and are related by SO(2) rotation). The fields $\kb$ and $\mb$ transform covariantly under \UI\ gauge transformations, however the term $\frac{\rho^2}{4}\left((\Tr{\kb\partial_i \lm})^2-(\Tr{\mb\partial_i \lm})^2 \right)\tttt_3+\frac{\rho^2}{2} \Tr{\kb\partial_i\lm}\Tr{\mb\partial_i\lm} \tttt_1$ is \UI\ (and \UC) gauge invariant. This is in contrast with the variables $\textbf{\textit{k}}$ and $\textbf{\textit{m}}$ from \cite{Faddeev_Niemi_SU2}, which are \UI\ (and \UC) gauge invariant. However, note that the field $\textbf{\textit{k}}$ and the parameter $\eta$ from \cite{Faddeev_Niemi_SU2} are not uniquely defined in the static case (it is the case $\mathbf{p}=0$ in \cite{Faddeev_Niemi_SU2}, the equation (28) in \cite{Faddeev_Niemi_SU2} is singular in this case and  $\textbf{\textit{k}}$ is not uniquely defined by equation (34) in \cite{Faddeev_Niemi_SU2}; neither the parameter $\eta$ is). This is caused by the fact, that in the static case the relevant Grassmanian is $\widetilde{\rm G}$(3,2)=SO(3)/SO(2) and not $\widetilde{\rm G}$(4,2)=SO(4)/(SO(2)$\times$SO(2)), and the parameter $\eta$ is spurious. We removed the extra parameter by choosing $\textbf{\textit{k}}=\kb$ and $\eta=0$, and declaring $\eta$ to be \UI\ gauge invariant, and $\kb$ to be \UI\ gauge covariant.  As a result, in the formula \eqref{eq:sigma} 
the field $\tttt$ appears instead of $\textbf{\textit{n}}$ from \cite{Faddeev_Niemi_SU2} (let us remind, that if $\eta=0$, then  $\textbf{\textit{n}}=\ttt$). The field $\tttt$ we use is related to $\ttt$ from \cite{Faddeev_Niemi_SU2} by a global rotation (in particular $\tttt_2=\ttt_3$). We show this in the next subsection.\footnote{There is also a difference in some coefficients. This is due to some typos in \cite{Faddeev_Niemi_SU2}. The coefficient in front of $(\partial_i \lv)^2$ is different, because there is a typos in equation (54) in \cite{Faddeev_Niemi_SU2}: there is $\rho^2|(\partial_a +i C_a) \mathbf{e}|=\frac{\rho^2}{2} ((\partial_a \mathbf{p})^2+(\partial_a \mathbf{q})^2)$ and there should be $\rho^2|(\partial_a +i C_a)\mathbf{e}|^2=2\rho^2 ((\partial_a \mathbf{p})^2+(\partial_a \mathbf{q})^2)$. The coefficient in front of $\epsilon_{ijk}l_k$ is different because there is a typos in first equation in section IX: there is $H_{ij}=\frac{1}{2\sqrt{2}}\epsilon_{ijk}l_k$ and there should be 
$H_{ij}=\frac{1}{2}\epsilon_{ijk}l_k$. The coefficient in 
front of $\lv\cdot (\partial_i \lv \times \partial_j \lv)$ is different, because there is a typos in second equation in section IX: there is $\partial_i \hat{C}_j-\partial_j \hat{C}_i=-\frac{1}{2\sqrt{2}}\lv\cdot (\partial_i \lv \times \partial_j \lv)$ and there should be $\partial_i \hat{C}_j-\partial_j \hat{C}_i=-\lv\cdot (\partial_i \lv \times \partial_j \lv)$.}
\subsection{The original Faddeev-Niemi approach}\label{sc:original_FN}
Although the decomposition \eqref{eq:decomposition_SU2} is equivalent to the original Faddeev-Niemi decomposition, it is not the same. The correspondence with the Faddeev-Niemi variables $\psi_1,\,\psi_2,\,e_i,\,\overline{e}_i$ is given by an U(3) matrix
$$
\omicron=\left(\begin{array}{ccc} \frac{1}{\sqrt{2}} & \frac{i}{\sqrt{2}} & 0\\ \frac{1}{\sqrt{2}} & -\frac{i}{\sqrt{2}} &0 \\ 0&0&1  \end{array} \right),
$$
in the following way:
$$
X^+= \psi\, \SOO, 
$$
where $\psi=[\psi_1, \psi_2 ,0]= \Phi \omicron^\dagger$, and $\SOO:=\omicron\, \SO$. The rows of the matrix $\SOO$ are $e:=\frac{1}{\sqrt{2}}\left(e^1+i\, e^2\right)$, $\overline{e}:=\frac{1}{\sqrt{2}}\left(e^1-i\, e^2\right)$ and $e^1\times e^2$. 
Therefore, the decomposition takes the following form:
$$
X^+_j= \psi_1 e_j + \psi_2 \overline{e}_j.
$$
Since
$$
\omicron \G  =  \left(\begin{array}{ccc} e^{i\lambda} & 0&0\\ 0& e^{-i \lambda} &0\\0&0&1\end{array} \right) \omicron,
$$
under the \UI\ gauge transformations $\psi$ transforms in the following way:
$$
\psi_1 \mapsto \psi_1 e^{i \lambda}, \quad \psi_2 \mapsto \psi_2 e^{-i \lambda},
$$
and $e_i$ transforms as
$$
e_i \mapsto e^{-i \lambda} e_i.
$$
Those are the transformation laws from \cite{Faddeev_Niemi_SU2}.

We introduce SU(2) matrix R:
$$
R=\frac{1}{\sqrt{|\psi_1|^2+|\psi_2|^2}}\left(\begin{array}{cc} \psi_1 & -\overline{\psi}_2\\ \psi_2& \overline{\psi}_1\end{array} \right).
$$

The chiral field $\ttt$ is:
\begin{equation}
\ttt:=\frac{1}{\sqrt{2}}R \sigma_3 R^\dagger.
\label{eq:sigma12}
\end{equation}
The coordinates of $\ttt$ read:
$$
t_i = \frac{1}{\sqrt{2}}\Tr{\sigma_i \ttt}
$$
It is an easy calculation to show, that a vector field $\ttt$ with coordinates $t_i$ coincides with the $\ttt$ field in \cite{Faddeev_Niemi_SU2}, i.e.
\begin{equation}
t_i= \frac{1}{\rho^2}\,[\overline{\psi}_1,\overline{\psi}_2]\,\sigma_i \left[\begin{array}{c}
                                                                  \psi_1\\ \psi_2
                                                                 \end{array}
 \right].\label{eq:ttt}
\end{equation}
Using equation \eqref{eq:ttt} and the fact, that $\psi=\Phi \omicron^\dagger$ it is easy to shown, that
$$
t_1= \tttt_3,\quad t_2=-\tttt_1,\quad t_3=\tttt_2.
$$
Therefore the $\tttt$ field we use and the $\ttt$ field from the original Faddeev-Niemi formulation differ only by a global O(3) transformation. 

Let us recall, that the chiral field $\lv$, as well as the gauge potential $C_i$ coincide in both representations. The only difference is in the fields $\kb$ and $\mb$, which in general are not equal to $\textbf{\textit{k}}$ and $\textbf{\textit{m}}$ from \cite{Faddeev_Niemi_SU2}. However, at each point they lie in the same plane and are related by an SO(2) rotation.

\section{The static SU(3) variables}\label{sc:variables}
In this section we present new parametrization of the static SU(3) connection motivated by the Faddeev-Niemi decomposition of the static SU(2) connection. To emphasize the similarity between those decompositions, and to make the comparison easier, the structure of this section resembles the structure of the previous one.
\subsection{Notation}\label{ssec:notation}
We start with introducing a notation we use in this section. Let $\lambda_a,\ a\in\{1,\ldots,8\}$ be the Gell-Mann matrices. We choose the basis of the su(3) algebra of the form:
\[
 \kappa_a=-\frac{i}{\sqrt{2}} \lambda_a.
\]
It has the property, that it is normalised in the scalar product:
\begin{equation}\label{eq:scalar_product}
 (v , w):={\rm Tr}\left(v^\dagger\, w\right),\ v,w\in su(3)\otimes\mathbb{C}.
\end{equation}
As a result for $v=v_a \kappa_a,\ w=w_a \kappa_a$
\[
 (v, w)= \overline{v_a} w_a.
\]
Repeated indices are summed. 

We denote by $f^{abc},\ a,b,c\in\{1,\ldots,8\}$ the su(3) structure constants, i.e.
\[
 [\kappa_a,\kappa_b]=f^{abc} \kappa_c.
\]
The structure constants are completely antisymmetric and take values:
\[
     f^{123} = \sqrt{2} \ , \quad f^{147} = f^{165} = f^{246} = f^{257} = f^{345} = f^{376} = \frac{\sqrt{2}}{2} \ , \quad f^{458} = f^{678} = \frac{\sqrt{6}}{2} \ .  
\]
We will use in this article the following (Cartan-Weyl) basis:
\[
 \kappa_1^{\pm}:=\frac{1}{\sqrt{2}}(\kappa_1\pm i\kappa_2),\quad \kappa_2^{\pm}:=\frac{1}{\sqrt{2}}(\kappa_4\mp i\kappa_5),\quad \kappa_3^{\pm}:=\frac{1}{\sqrt{2}}(\kappa_6\pm i\kappa_7),\quad \kappa_3,\quad \kappa_8.
\]
It is an orthonormal basis in the scalar product \eqref{eq:scalar_product}. 
\subsection{Singular value decomposition of the connection}\label{ssec:decomposition}
In analogy to the SU(2) case we express the gauge potential $A_\mu$ in the Cartan-Weyl basis
\begin{equation}\label{basis}
 A_\mu={B}^{1}_\mu \kappa_1^+ + B^2_\mu \kappa_2^++B^3_\mu \kappa_3^++\overline{B^1_\mu} \kappa_1^- + \overline{B^2_\mu} \kappa_2^- +\overline{B^3_\mu} \kappa_3^- + A_\mu^3 \kappa_3 + A_\mu^8 \kappa_8,\ \mu\in\{0,1,2,3\}.
\end{equation}

In the static limit (i.e. $\partial_0 A_\mu \equiv 0,\ A_0\equiv 0$) the $B$ field may be considered to be a 3x3 matrix (in this article we assume that $B$ is invertible). We write the matrix $B\indices{^{i}_j}(x),\ i,j\in\{1,2,3\}$ using the singular value decomposition:
\begin{equation}\label{eq:decomposition}
 B=U\, \rho\, V^\dagger,
\end{equation}
where $U(x)$ and $V(x)$ are 3x3 unitary matrices and $\rho(x)$ is diagonal matrix with positive eigenvalues $\rho_1,\,\rho_2,\,\rho_3$. The decomposition is not unique. We partially fix the ambiguity by requiring that $V$ is an SU(3) matrix. The matrix $U$ is a U(3) matrix, and its determinant can be arbitrary complex number with unit norm. We define:
$$
e^{3\iu \phi}:=\det (U).
$$

We do not fix the ambiguity completely, there is always U(1)$\times$U(1) symmetry of the decomposition. A comparison of the formula \eqref{eq:decomposition} with formula \eqref{eq:decomposition_SU2} shows strong structural similarity between this decomposition and Faddeev-Niemi decomposition. This analogy is fundamental for the considerations in this section: thanks to the structural similarity, the definition of chiral fields, internal gauge potentials and supercurrents is easily transferred from the previous section; also many of the calculations are similar.
\subsubsection{An interpretation of the variables}\label{sec:interpretation}
Before going further, let us first present an interpretation of the decomposition, which makes it easier to see analogies between the decomposition we introduce for SU(3) connection and the original Faddeev-Niemi decomposition of the SU(2) connection. 

The squares of the fields $\rho_1,\,\rho_2,\,\rho_3$ are eigenvalues of a matrix:
\begin{equation}\label{eq:hmatrix}
h^{pq}:= B^p_i \overline{B^q_i}. 
\end{equation}
It is in fact hermitian and positive semi-definite. The fields  $\rho_1,\,\rho_2,\,\rho_3$ are therefore analogous to the scalar field $\rho$ in the SU(2) case (which is an eigenvalue of a 1$\times$1 matrix $X^+_i \overline{X^+_i}$). The matrix $h(x)$ can be diagonalized with an U(3) matrix 
$$
h = U \rho^2 U^{\dagger}.
$$
We introduce auxiliary variables -- vectors $v^1_i, v^2_i, v^3_i$ such that
\begin{equation}\label{eia}
 v^p_i:= U\indices{^\dagger^p_q} B\indices{^q_i}.
\end{equation}
They satisfy (we do not sum over $p$ in this formula):
$$
 v^p_i \overline{v^q_i}= \rho^2_p\, \delta^{pq}.
$$
This means, that they are orthogonal in the scalar product
\[
 (A|B):=(A_i, B_i).
\]
In this paper we focus on the case when $B$ is invertible, i.e.
$$\rho_1\not=0,\,\rho_2\not=0,\,\rho_3\not=0.$$
As a result none of the vectors $v^1_i, v^2_i, v^3_i$ is a zero vector, and $v^p_i$ form a basis of $\mathbb{C}^3$. We further normalize the basis, i.e. we introduce three complex vector fields
\[
 e^1_i:= \frac{1}{\rho_1} v^1_i,\quad e^2_i:= \frac{1}{\rho_2} v^2_i,\quad e^3_i:= \frac{1}{\rho_3} v^3_i.
\]
They are orthonormal and we consider them to be analogous to the zweibein $e_i, \overline{e}_i$. The matrix $U$ is not unique, neither are the vectors $e^p_i$. The ambiguity can be partially fixed by requiring that those vectors form an SU(3) matrix $V$:
\begin{equation}
 (V^\dagger)^p_q = e^p_q,\quad p,q\in\{1,2,3\}.
\end{equation}
\subsection{Gauge transformations}\label{sssec:ambiguity}
\subsubsection{External gauge transformations}
 There is a U(1)$\times$U(1) symmetry of the Lagrangian which is the symmetry remaining after Maximal Abelian (partial) Gauge fixing. We call it external gauge symmetry and denote by \UCUC. The corresponding covariant derivative acts on $U$ in the following way:
\begin{eqnarray*}D_i U:= \partial_i U + \sqrt{3}\left(\frac{\sqrt{3}}{2} A^3_i + \frac{1}{2} A^8_i \right) \kappa_3 U + \sqrt{3}\left(\frac{1}{2} A^3_i - \frac{\sqrt{3}}{2} A^8_i \right) \kappa_8 U, \\
 D_i U^\dagger:= \partial_i U^\dagger - \sqrt{3}\left(\frac{\sqrt{3}}{2} A^3_i + \frac{1}{2} A^8_i \right) U^\dagger \kappa_3 - \sqrt{3}\left(\frac{1}{2} A^3_i - \frac{\sqrt{3}}{2} A^8_i \right) U^\dagger \kappa_8 .
\end{eqnarray*}
\subsubsection{Internal gauge transformations}
As we noted before, the decomposition \eqref{eq:decomposition} is not unique. There is a U(1)$\times$U(1) symmetry of this decomposition:
\begin{equation}\label{eq:gauge}
 U\mapsto U G,\quad V \mapsto V G,
\end{equation}
where $G=e^{-i\omega^3 \lambda_3-i\omega^8 \lambda_8}$ takes values in the diagonal U(1)$\times$U(1) subgroup of SU(3).
We call it an internal symmetry and denote by \UIUI. We define a \UIUI\ gauge potential $\C_i = \C_i^3 \kappa_3^\dagger + \C_i^8 \kappa_8^\dagger$:
\begin{equation}\label{eq:C}
\C_i^3=\Tr{\kappa_3 V^\dagger \partial_i V},\quad  \C_i^8=\Tr{\kappa_8 V^\dagger \partial_i V}.
\end{equation}
This gauge potential is analogous to the \UI\ gauge potential $C_i$ \eqref{eq:Ci}.
We introduce a derivative:
\begin{eqnarray*}
\DCC{i} U = D_i U - U \C_i , \quad \DCC{i} U^\dagger = D_i U^\dagger + \C_i U^\dagger.
\end{eqnarray*}
It is covariant with respect to \UCUC\ and \UIUI\ gauge transformation as well as global shifts in the phase of $U$.

\subsection{Decomposition of $\rho^2$}\label{sc:rho2}
A trace part of the matrix $\rho^2$ has an interpretation of a condensate. We therefore extract this part by introducing new variables:
\begin{eqnarray}
 &\chi^2:=\frac{1}{6}\Tr{\rho^2}=\frac{1}{6}\left(\rho_1^2+\rho_2^2+\rho_3^2\right)\label{eq:chi2}\\
 &\x:=-\frac{\iu}{2\sqrt{6}}\Tr{\kappa_3^\dagger \rho^2}=\frac{1}{4\sqrt{3}} \left(\rho_1^2-\rho_2^2 \right)\label{eq:u}\\
 &\y:=-\frac{\iu}{2\sqrt{6}}\Tr{\kappa_8^\dagger \rho^2}=\frac{1}{12} \left(\rho_1^2+\rho_2^2-2\rho_3^2\right).\label{eq:v}
\end{eqnarray}
Since the relations can be inverted, those variables replace the (positive) scalars $\rho_1,\ \rho_2,\ \rho_3$. We also introduce polar counterpart of the coordinates $\chi^2, r ,\alpha$, where
\begin{equation}\label{eq:alpha}
\x = -\, r \sin(2\alpha), \quad \y= r \cos(2\alpha).
\end{equation}
We will think of the functions $\rho_1,\, \rho_2,\, \rho_3$ as functions determined by $\chi^2:\mathbb{R}^3\to \mathbb{R}_+$, $r:\mathbb{R}^3\to \mathbb{R}_+$, $\alpha:\mathbb{R}^3\to \mathbb{R}\,{\rm mod}\,\pi$ (in this interpretation we assume, that $\rho_1^2,\, \rho_2^2,\, \rho_3^2$  determined by the equations \eqref{eq:chi2}, \eqref{eq:u}, \eqref{eq:v} are positive numbers; this imposes constraints on $\chi^2,\, r,\,\alpha$, and we assume, that $\chi^2,\, r,\,\alpha$ satisfy these constraints).
\subsection{Chiral fields and supercurrents}
\subsubsection{Chiral fields}
The chiral fields corresponding to $U$ are the following:
\begin{equation}\label{eq:nm}
 \n:=U^\dagger \kappa_3 U, \quad \m:=U^\dagger \kappa_8 U.
\end{equation}
Analogously to the SU(2) case, the fields $\n$ and $\m$ are \UCUC\ invariant. The covariant derivative $\DCC{i}$ naturally transfers to the fields $\n$ and $\m$:
$$
\DCC{i}\n=\partial_i \n + [\C_i,n],\quad \DCC{i}\m=\partial_i \m +[\C_i,\m].
$$
The chiral fields corresponding to $V$ are the following:
\begin{equation}\label{eq:nnmm}
\nn:= V\kappa_3^\alpha V^\dagger,\quad \mm:=V\kappa_8^\alpha V^\dagger,
\end{equation}
where
$$
\kappa_3^\alpha:=\cos(\alpha) \kappa_3 - \sin(\alpha)\kappa_8, \quad \kappa_8^\alpha:=\sin(\alpha) \kappa_3 + \cos(\alpha)\kappa_8.
$$
The $\alpha$ variable is defined in \eqref{eq:alpha}. Importantly, note that the structural difference between \eqref{eq:nnmm} and \eqref{eq:nm} is not only in the $\alpha$ parameter but also in the way daggers are placed. As a result the fields $\nn$ and $\mm$ are \UIUI\  invariant. This mimics the \UI\ invariance of the $\lm$ field \eqref{eq:l}.
\subsubsection{Supercurrents}
The supercurrents are defined by the following equations (compare with \eqref{eq:supercurrentSU2}):
\begin{equation}\label{eq:supercurrent}
\sJ{i}{3}:= \Tr{\kappa_3 U \DCC{i} U^\dagger}, \quad \sJ{i}{8}:= \Tr{\kappa_8 U \DCC{i} U^\dagger}.
\end{equation}
\subsection{The static SU(3) Yang-Mills Lagrangian}
As in the SU(2) case, we partially fix the gauge using Maximal Abelian Gauge fixing $D^{\mu} B_{\mu}^p=0, p\in\{1,2,3\}$, where the covariant derivative is:
\begin{eqnarray*}
 D_\mu B^1_\nu:=\partial_\mu B^1_\nu-i\sqrt{2}A^3_\mu B^1_\nu,\ D_{\mu} B^2_\nu:=\partial_\mu B^2_\nu+i\frac{\sqrt{2}}{2}A^3_\mu B^2_\nu+i\frac{\sqrt{6}}{2}A^8_\mu B^2_\nu,\ D_\mu B^3_\nu:=\partial_\mu B^3_\nu+i\frac{\sqrt{2}}{2}A^3_\mu B^3_\nu-i\frac{\sqrt{6}}{2}A^8_\mu B^3_\nu,\\
 D_\mu \overline{B^1_\nu}:=\partial_\mu\overline{B^1_\nu}+i\sqrt{2}A^3_\mu \overline{B^1_\nu},\ D_\mu \overline{B^2_\nu}:=\partial_\mu\overline{B^2_\nu}-i\frac{\sqrt{2}}{2}A^3_\mu \overline{B^2_\nu}-i\frac{\sqrt{6}}{2}A^8_\mu \overline{B^2_\nu},\ D_\mu\overline{B^3_\nu}:=\partial_\mu\overline{B^3_\nu}-i\frac{\sqrt{2}}{2}A^3_\mu \overline{B^3_\nu}+i\frac{\sqrt{6}}{2}A^8_\mu \overline{B^3_\nu}.
\end{eqnarray*}
We introduce the following tensors :
\begin{eqnarray*}
 &\Pt_{\mu\nu}=-\iu  \sqrt{2} B^1_{[\mu} \overline{B^1_{\nu]}}+\iu\frac{\sqrt{2}}{2} B^2_{[\mu} \overline{B^2_{\nu]}}+\iu\frac{\sqrt{2}}{2} B^3_{[\mu}\overline{B^3_{\nu]}},\\
&\Pe_{\mu\nu}=\iu\frac{\sqrt{6}}{2} B^2_{[\mu} \overline{B^2_{\nu]}}-\iu\frac{\sqrt{6}}{2} B^3_{[\mu}\overline{B^3_{\nu]}}.
\end{eqnarray*}
The bracket denotes here the antisymmetrization of indices: $A_{[\mu \nu]}=A_{\mu\nu}-A_{\nu\mu}$. The tensors  $\Pt$ and $\Pe$ are analogous to the tensor $P$ in the SU(2) case \eqref{eq:YM_action}. 

We study the static limit. In this case, the Lagrangian explicitly reads (see \cite{Bolokhov_Faddeev} and Appendix \ref{sc:YMCW}):
\begin{equation}\label{eq:lagrangian}
L_{\rm SU(3)}=\frac{1}{2} |D_{i} B_{j}^p +\iu \epsilon^{pqr} \overline{B^q_i B^r_j} |^2- \Tr{h}^2+ \Tr{h^2} +\frac{1}{4} \left(\partial_{[i} A^3_{j]}+2\Pt_{ij} \right)^2+\frac{1}{4} \left(\partial_{[i} A^8_{j]}+2\Pe_{ij} \right)^2-\frac{3}{4}(\Pt_{ij})^2-\frac{3}{4}(\Pe_{ij})^2.
\end{equation}
Let us recall, that the matrix $h$ was defined in formula \eqref{eq:hmatrix}. It satisfies  $\Tr{h}=\Tr{\rho^2},\, \Tr{h^2}=\Tr{\rho^4}$.

\subsection{The term $\frac{1}{2} |D_{i} B_{j}^p +\iu \epsilon^{pqr} \overline{B^q_i B^r_j} |^2$}
As in the SU(2) case, the Lagrangian naturally splits into two parts. The first part is  $\frac{1}{2} |D_{i} B_{j}^p +\iu \epsilon^{pqr} \overline{B^q_i B^r_j} |^2$. It contains the standard sigma model Hamiltonians of the chiral fields.
\subsubsection{The term $\iu \epsilon^{pqr} \overline{B^q_i B^r_j}$}
A comparison of the SU(3) static Lagrangian \eqref{eq:lagrangian} with the SU(2) static Lagrangian \eqref{eq:YM_action} reveals a structural difference between them. The difference is in the term $\iu \epsilon^{pqr} \overline{B^q_i B^r_j}$. As a result, in the final expression there will appear terms that can not be compared to any terms in the SU(2) case. 

We interpret the term $\iu \epsilon^{pqr} \overline{B^q_i B^r_j}$ using the matrix interpretation of $B(x)$. Let us recall, that in this article we consider non-degenerate case, i.e. $B(x)$ is invertible. This allows us to write:
$$
\iu \epsilon^{pqr} \overline{B^q_i B^r_j}=\iu \epsilon^{sqr} \overline{B^s_k B^q_i B^r_j} (\overline{B}^{-1})^k_p=\iu \det (\overline{B}) \epsilon_{kij} (\overline{B}^{-1})^k_p.
$$
Let us recall, that we introduced three matrices \eqref{eq:epsilon}:
$$
(\epsilon_k)_{ij}:= \epsilon_{kij}, \quad k,i,j\in\{1,2,3\}.
$$
Note, that they may be interpreted in terms of Gell-Mann matrices:
$$\epsilon_{1ij}=-\sqrt{2}\,(\kappa_7)_{ij},\quad \epsilon_{2ij}=\sqrt{2}\,(\kappa_5)_{ij},\quad \epsilon_{3ij}=-\sqrt{2}\,(\kappa_2)_{ij}.$$ 
Using them, we can write:
$$
\iu \epsilon^{pqr} \overline{B^q_i B^r_j}=-\iu \det (\overline{B}) \left((\overline{B}^{-1})^* \epsilon_i\right)^p_j,
$$
where ${}^*$ denotes the matrix transposition.
\subsubsection{Applying singular value decomposition}
We apply now the singular value decomposition of the gauge potential:
$$
\iu \epsilon^{pqr} \overline{B^q_i B^r_j}=-\iu e^{-3\iu\phi} \det(\rho) \left(U \rho^{-1} V^\dagger \epsilon_i\right)^p_j.
$$

The term $\frac{1}{2} |D_{i} B_{j}^p +\iu \epsilon^{pqr} \overline{B^q_i B^r_j} |^2$ obtains the following matrix representation:
$$\frac{1}{2} \lVert D_{i} \left( U \rho V^\dagger\right)-\iu e^{-3\iu\phi} \det(\rho) U \rho^{-1} V^\dagger \epsilon_i \rVert^2,$$
where $\lVert \cdot \rVert$ denotes the Hilbert-Schmidt norm $\lVert M \rVert^2:=\Tr{M^\dagger M}$.

Expanding the expression, we obtain:
\begin{eqnarray*}
 &\frac{1}{2} |D_{i} B_{j}^p +\iu \epsilon^{pqr} \overline{B^q_i B^r_j} |^2=\frac{1}{2}\Tr{D_i U \rho^2 D_i U^\dagger}+\frac{1}{2} \Tr{\partial_i \rho \partial_i \rho} + \frac{1}{2} \Tr{D_i V \rho^2 D_i V^\dagger} + \det (\rho)^2 \Tr{\rho^{-2}}+\\&+\Tr{U^\dagger D_i U \rho V^\dagger \partial_i V \rho}-\frac{\iu}{2} e^{3\iu \phi} \det (\rho)\Tr{U^\dagger D_i U \rho V^\dagger \epsilon_i V \rho^{-1}}+\frac{\iu}{2} e^{-3\iu \phi} \det (\rho)\Tr{U^\dagger D_i U \rho^{-1} V^\dagger \epsilon_i V \rho}-\\&-\iu \cos(3 \phi) \det(\rho) \Tr{\rho^{-1}\partial_i \rho V^\dagger \epsilon_i V}+\sin(3\phi) \det(\rho) \Tr{V\partial_i V^\dagger \epsilon_i}.
\end{eqnarray*}
As in the SU(2) case (see section \ref{sc:eginv_form}) we will write this expression as a sum of terms which are \UIUI\ gauge invariant. We call the resulting expression an explicitly gauge invariant form.
\subsubsection{Explicitly gauge invariant form}
Let $\Proj: su(3) \to su(3)$ be the orthogonal projector onto a space orthogonal to the Cartan subalgebra spanned by $\kappa_3$ and $\kappa_8$:
$$
\Proj (v) = v - (v,\kappa_3) \kappa_3 - (v,\kappa_8) \kappa_8.
$$
Importantly, note that $\Proj (V^\dagger \partial_i V)$ is \UIUI\ gauge covariant:
$$
\Proj(V^\dagger \partial_i V) \mapsto G^\dagger \Proj(V^\dagger \partial_i V) G.
$$
The projector $\Proj$ is analogous to the projector $\proj$ in the SU(2) case (see section \ref{sc:eginv_form}). Using the projector $\Proj$ and covariant derivatives introduced in the section \ref{sssec:ambiguity}, one can write the term $ \frac{1}{2} |D_{i} B_{j}^p +\iu \epsilon^{pqr} \overline{B^q_i B^r_j} |^2$ in a form which is explicitly \UCUC\ and \UIUI\ invariant (i.e. each term is \UCUC\ and \UIUI\ gauge invariant):
\begin{eqnarray*}
 &\frac{1}{2} |D_{i} B_{j}^p +\iu \epsilon^{pqr} \overline{B^q_i B^r_j} |^2=-\frac{1}{2}\Tr{U^\dagger \DCC{i} U \rho^2  U^\dagger \DCC{i} U}+\Tr{U^\dagger \DCC{i} U \rho \Proj( V^\dagger \partial_i V )\rho}- \frac{1}{2} \Tr{\Proj(V^\dagger \partial_i V) \rho^2 \Proj(V^\dagger \partial_i V)}-\\&-\frac{\iu}{2} e^{3\iu \phi} \det (\rho)\Tr{U^\dagger \DCC{i} U \rho V^\dagger \epsilon_i V \rho^{-1}}+\frac{\iu}{2} e^{-3\iu \phi} \det (\rho)\Tr{U^\dagger \DCC{i} U \rho^{-1} V^\dagger \epsilon_i V \rho}-\\&-\iu \cos(3 \phi) \det(\rho) \Tr{\rho^{-1}\partial_i \rho V^\dagger \epsilon_i V}-\sin(3\phi) \det(\rho) \Tr{\Proj(V^\dagger\partial_i V) V^\dagger \epsilon_i V}+\frac{1}{2} \Tr{\partial_i \rho \partial_i \rho} + \det (\rho)^2 \Tr{\rho^{-2}}.
\end{eqnarray*}
\subsubsection{The term $-\frac{1}{2}\Tr{U^\dagger \DCC{i} U \rho^2  U^\dagger \DCC{i} U}$}\label{sc:DUDU}
Note, that $\frac{1}{2} |D_{i} B_{j}^p +\iu \epsilon^{pqr} \overline{B^q_i B^r_j} |^2$ depends on $U$ via the one-form $U^\dagger \DCC{i} U$. Using a trick similar to the one which lead to equation \eqref{eq:QdQtDt} in the SU(2) case, we are able to express this one-form in terms of chiral fields and supercurrents. 
We decompose $U^\dagger \DCC{i} U$ into the part parallel to the Cartan subalgebra spanned by $\n$ and $\m$, and the part orthogonal to the subalgebra. This decomposition is based on the following observation \cite{Faddeev_Niemi_SUN}: the operator $P:{\rm u(3)}\to {\rm u(3)}$, defined by
\begin{equation}\label{eq:projection}Pv:=-\frac{1}{2}([\n,[\n,v]]+ [\m,[\m,v]]),\quad v\in{\rm u(3)}\end{equation}
 is the orthogonal projection operator onto the subspace of u(3) that is orthogonal to the Cartan subalgebra spanned by $\n,\,\m$ and $\mathbbm{1}$ (the identity matrix). We have
$$P(U^\dagger \DCC{i} U)= -\frac{1}{2}([\n, [\n, U^\dagger \DCC{i} U]] + [\m, [\m, U^\dagger \DCC{i} U]) $$
Note, that
$$
\DCC{i} \n = [\n, U^\dagger \DCC{i} U],\quad \DCC{i} \m = [\m, U^\dagger \DCC{i} U].
$$
Therefore
$$P(U^\dagger \DCC{i} U)=-\frac{1}{2}([\n, \DCC{i} \n] + [\m, \DCC{i} \m]). $$
On the other hand, from the definition of $P$ follows, that
$$
P(U^\dagger \DCC{i} U)= U^\dagger \DCC{i} U - (\n,U^\dagger \DCC{i} U)\,\n-(\m,U^\dagger \DCC{i} U)\,\m - \frac{1}{3}\Tr{U^\dagger \DCC{i} U} \mathbbm{1}.
$$
As a result, the decomposition takes the following form (see also \cite{Shabanov_UdU}):
\begin{equation}\label{eq:UdU}
 U^\dagger \DCC{i} U= -\frac{1}{2}([\n, \DCC{i} \n] + [\m, \DCC{i} \m]) + \sJ{i}{3}\, \n+ \sJ{i}{8}\, \m+\iu \partial_i \phi\, \mathbbm{1}.
\end{equation}
Let us recall, that $\sJ{i}{3}$ and $\sJ{i}{8}$ are the supercurrents defined in \eqref{eq:supercurrent}.
Using the variables $\chi^2,r,\alpha$ from section \ref{sc:rho2} and formula \eqref{eq:UdU} one can show that:
\begin{eqnarray}
-\frac{1}{2}\Tr{U^\dagger \DCC{i} U \rho^2  U^\dagger \DCC{i} U}=\frac{1}{2}\chi^2 \left( \DCC{i} \n, \DCC{i}\n \right)+\frac{1}{2}\chi^2 \left( \DCC{i} \m, \DCC{i}\m \right)+\chi^2(\sJ{i}{3})^2+\chi^2(\sJ{i}{8})^2+3\chi^2 (\partial_i \phi)^2-\label{eq:sigma1}\\
-\iu \sqrt{6}\,r\,\Tr{\left(-\frac{1}{2}([\n, \DCC{i} \n] + [\m, \DCC{i} \m]) + \sJ{i}{3}\, \n+ \sJ{i}{8}\, \m+\iu \partial_i \phi\right)^2 \kappa_8^{-2\alpha}},
\end{eqnarray}
where $\kappa_8^{-2\alpha}=\sin(-2\alpha)\,\kappa_3+\cos(-2\alpha)\,\kappa_8$. The terms that depend on $\chi^2$ contain: the \textbf{standard sigma model Hamiltonian} expected in \cite{Faddeev_Niemi_SUN}, terms which make the supercurrents $\sJ{i}{{3/8}}$ massive, and a term quadratic in derivatives of $\phi$. This equation is analogous to the equation \eqref{eq:DCQDCQ} in the SU(2) case.
\subsubsection{The terms $ - \frac{1}{2} \Tr{\Proj(V^\dagger \partial_i V) \rho^2 \Proj(V^\dagger \partial_i V)}$ and $-\sin(3\phi) \det(\rho) \Tr{\Proj(V^\dagger\partial_i V) V^\dagger \epsilon_i V}$}
We have chosen to analyse those two terms in the same section, because they both share the property that they depend on $\Proj(V^\dagger \partial_i V)$ and do not depend on $U$, and similar technique may be used in the analysis. We will perform an analysis of the left-invariant form $V^\dagger \partial_i V$ similar to the analysis of the form $U^\dagger \DCC{i} U$ presented in the previous subsection, and we will express those terms using chiral fields $\nn$ and $\mm$. Note, that there is a subtle difference in expressing the form $U^\dagger \DCC{i} U$ in terms of chiral fields and supercurrents, and expressing the form $\Proj(V^\dagger \partial_i V)$ in terms of chiral fields. The difference is in the placements of daggers in the definition of $\nn, \mm$ fields and $\n, \m$. 

Let us start with the term
\begin{equation*}
 - \frac{1}{2} \Tr{\Proj(V^\dagger \partial_i V) \rho^2 \Proj(V^\dagger \partial_i V)}
\end{equation*}
As one may expect after reading the previous subsection, the term will give rise to standard sigma model Hamiltonian terms. 

We decompose $V^\dagger \partial_i V$ in the Cartan-Weyl basis of the su(3) Lie algebra:
\begin{equation*}
 {\cal J}_i^p \kappa_p^++\overline{{\cal J}_i^p} \kappa_p^--\C_i^3\kappa_3-\C_i^8\kappa_8:=V^\dagger \partial_i V.
\end{equation*}
It is easy to verify, that
\begin{equation}\label{eq:rhokappa}
 \rho^2 \kappa^+_1 = \rho_1^2\, \kappa^+_1,\quad  \rho^2 \kappa^-_1 = \rho_2^2\, \kappa^-_1,\quad\rho^2 \kappa^+_2 = \rho_3^2\, \kappa^+_2,\quad  \rho^2 \kappa^-_2 = \rho_1^2\, \kappa^-_2,\quad \rho^2 \kappa^+_3 = \rho_2^2\, \kappa^+_3,\quad  \rho^2 \kappa^-_3 = \rho_3^2\, \kappa^-_3.
\end{equation}
It follows now, that
\begin{equation}\label{eq:VdVrVdV}
 -\Tr{\Proj(V^\dagger \partial_i V) \rho^2 \Proj(V^\dagger \partial_i V)}=(\rho_1^2 +\rho_2^2)\,\overline{{\cal J}_i^1} {\cal J}_i^1+(\rho_1^2+\rho_3^2)\, \overline{{\cal J}_i^2} {\cal J}_i^2+(\rho_2^2+\rho_3^2)\, \overline{{\cal J}_i^3} {\cal J}_i^3.
\end{equation}
Written in this form, this term may be further interpreted using the fields $\nn$ and $\mm$.

One may easily show, that 
\begin{equation}\label{eq:dnndmm}
 \partial_i \nn=[\nn,V \partial_i V^\dagger]-\partial_i \alpha\,\mm ,\quad \partial_i \mm=[\mm,V \partial_i V^\dagger]+\partial_i \alpha\,\nn.
\end{equation}

We calculate 
\begin{eqnarray*}
& (\partial_i \nn,\partial_i \nn)-(\partial_i \alpha)^2=([\nn,V \partial_i V^\dagger],[\nn, V \partial_i V^\dagger])=-([\nn,[\nn,V \partial_i V^\dagger]],V \partial_i V^\dagger)=-([\kappa_3^\alpha,[\kappa_3^\alpha,V^\dagger \partial_i V]],V^\dagger \partial_i V)=\\
&=4 \cos^2 (\alpha)\, \overline{{\cal J}_i^1} {\cal J}_i^1+4 \cos^2 (\alpha+\frac{\pi}{3})\, \overline{{\cal J}_i^2} {\cal J}_i^2+4 \cos^2 (\alpha-\frac{\pi}{3})\, \overline{{\cal J}_i^3} {\cal J}_i^3.
\end{eqnarray*}
Similarly
\begin{equation*}
 (\partial_i \mm,\partial_i \mm)-(\partial_i \alpha)^2=4 \sin^2 (\alpha)\, \overline{{\cal J}_i^1} {\cal J}_i^1+4 \sin^2 (\alpha+\frac{\pi}{3})\, \overline{{\cal J}_i^2} {\cal J}_i^2+4 \sin^2 (\alpha-\frac{\pi}{3})\, \overline{{\cal J}_i^3} {\cal J}_i^3.
\end{equation*}
It will be convenient for us to use linear combinations of those terms:
\begin{eqnarray}
  &(\partial_i \nn,\partial_i \nn)+(\partial_i \mm,\partial_i \mm)-2(\partial_i \alpha)^2=4\, \overline{{\cal J}_i^1} {\cal J}_i^1+4\, \overline{{\cal J}_i^2} {\cal J}_i^2+4\, \overline{{\cal J}_i^3} {\cal J}_i^3,\label{eq:nnJJ3}\\
  &(\partial_i \nn,\partial_i \nn)-(\partial_i \mm,\partial_i \mm)=4 \cos (2\,\alpha)\, \overline{{\cal J}_i^1} {\cal J}_i^1+4 \cos (2\,\alpha+\frac{2\pi}{3})\, \overline{{\cal J}_i^2} {\cal J}_i^2+4 \cos (2\,\alpha-\frac{2\pi}{3})\, \overline{{\cal J}_i^3} {\cal J}_i^3.\label{eq:nnJJ8}
\end{eqnarray}
It turns out, that the term \eqref{eq:VdVrVdV} is a linear combination of the terms \eqref{eq:nnJJ3}, \eqref{eq:nnJJ8}. The coefficients are $\chi^2$ and $r$:
\begin{eqnarray}
 \chi^2\,\left((\partial_i \nn,\partial_i \nn)+(\partial_i \mm,\partial_i \mm)-2(\partial_i \alpha)^2\right)+ r\,\left((\partial_i \nn,\partial_i \nn)-(\partial_i \mm,\partial_i \mm) \right)=&\nonumber\\=(\rho_1^2 +\rho_2^2)\,\overline{{\cal J}_i^1} {\cal J}_i^1+(\rho_1^2+\rho_3^2)\, \overline{{\cal J}_i^2} {\cal J}_i^2&+(\rho_2^2+\rho_3^2)\, \overline{{\cal J}_i^3} {\cal J}_i^3.\label{eq:bg}
\end{eqnarray}

Summarizing, we have:
\begin{equation}\label{eq:sigma2}
-\frac{1}{2}\Tr{\Proj(V^\dagger \partial_i V) \rho^2 \Proj(V^\dagger \partial_i V)}=\frac{1}{2}\chi^2\,\left(\partial_i \nn,\partial_i \nn\right)+\frac{1}{2}\chi^2\,\left(\partial_i \mm,\partial_i \mm\right)+\frac{1}{2}r\left(\partial_i \nn,\partial_i \nn\right)-\frac{1}{2}r\left(\partial_i \mm,\partial_i \mm\right)-\chi^2 (\partial_i \alpha)^2.
\end{equation}
As in the subsection \ref{sc:DUDU}, the term that depends on $\chi^2$ is the \textbf{standard sigma model Hamiltonian} expected in \cite{Faddeev_Niemi_SUN}. This equation is therefore analogous to the equation \eqref{eq:dldl} in the SU(2) case.

Note, that $\chi^2>r$, and the term has also a different interpretation:
\begin{equation}\label{eq:sigmaext}
-\frac{1}{2}\Tr{\Proj(V^\dagger \partial_i V) \rho^2 \Proj(V^\dagger \partial_i V)}=\frac{1}{2}(\chi^2+r)\,\left(\partial_i \nn,\partial_i \nn\right)+\frac{1}{2}(\chi^2-r)\,\left(\partial_i \mm,\partial_i \mm\right)-\chi^2 (\partial_i \alpha)^2.
\end{equation}
This suggests, that also a generalization of the model \eqref{eq:action} may play an important role in knot-like scenario for SU(3) Yang-Mills theory. In this generalization, the vectors $\kappa_3$ and $\kappa_8$ are rotated by the angle $\alpha$, and the coefficient in front of $\left(\partial_i \nn,\partial_i \nn\right)$ is in general different from the coefficient in front of $\left(\partial_i \mm,\partial_i \mm\right)$. The extended Faddeev-Niemi model \eqref{eq:action} \cite{Faddeev_Niemi_SUN,Shabanov} is recovered when the coefficients are equal. The remaining parameter $\alpha$ describes then a hidden symmetry of the extended Faddeev-Niemi model. Note also that the Shabanov bound provides a lower bound for the generalized model \eqref{eq:sigmaext}: simply use the fact that all three terms on the RHS of \eqref{eq:bg} are positive and the coefficients $\rho_1^2+\rho_2^2$, $\rho_1^2+\rho_3^2$, $\rho_2^2+\rho_3^2$ are bounded from below by $\min(\rho_1^2+\rho_2^2, \rho_1^2+\rho_3^2, \rho_2^2+\rho_3^2)$. The 
generalization requires, that we 
interpret all three fields $\rho_1,\, \rho_2,\, \rho_3$ as condensates. In this article we focus on a scenario, when $\chi^2$ becomes a condensate and use the interpretation \eqref{eq:sigma2}.

Let us analyse now the second term. Note, that
$$
\Proj (V^\dagger\partial_i V)=-\frac{1}{2} [\kappa_3^\alpha,[\kappa_3^\alpha,V^\dagger\partial_i V]]-\frac{1}{2} [\kappa_8^\alpha,[\kappa_8^\alpha,V^\dagger\partial_i V]].
$$
As a result
$$
V \Proj (V^\dagger\partial_i V) V^\dagger=\frac{1}{2} [\nn,[\nn,V \partial_i V^\dagger]]+\frac{1}{2} [\mm,[\mm,V \partial_i V^\dagger]].
$$
From equation \eqref{eq:dnndmm} follows now, that
\begin{equation}\label{eq:VdV}
 V \Proj (V^\dagger\partial_i V) V^\dagger=\frac{1}{2} [\nn,\partial_i\nn]+\frac{1}{2} [\mm,\partial_i \mm].
\end{equation}
Finally, we obtain the following expression:
$$
-\sin(3\phi) \det(\rho) \Tr{\Proj(V^\dagger\partial_i V) V^\dagger \epsilon_i V}=\frac{1}{2}\sin(3\phi) \det(\rho)(\partial_i \nn, [\epsilon_i,\nn])+\frac{1}{2}\sin(3\phi) \det(\rho)(\partial_i \mm, [\epsilon_i,\mm]).
$$
There is no analogous term in the static SU(2) Yang-Mills Lagrangian.
\subsubsection{Interaction terms: $\Tr{U^\dagger \DCC{i} U \rho \Proj( V^\dagger \partial_i V )\rho}, -\frac{\iu}{2} e^{3\iu \phi} \det (\rho)\Tr{U^\dagger \DCC{i} U \rho V^\dagger \epsilon_i V \rho^{-1}}+ \frac{\iu}{2} e^{-3\iu \phi} \det (\rho)\Tr{U^\dagger \DCC{i} U \rho^{-1} V^\dagger \epsilon_i V \rho}$}
In this section we discuss terms which describe an interaction of the two nonlinear sigma models. Those terms depend on both $U$ and $V$:
$$\Tr{U^\dagger \DCC{i} U \rho \Proj( V^\dagger \partial_i V )\rho},\quad -\frac{\iu}{2} e^{3\iu \phi} \det (\rho)\Tr{U^\dagger \DCC{i} U \rho V^\dagger \epsilon_i V \rho^{-1}}+ \frac{\iu}{2} e^{-3\iu \phi} \det (\rho)\Tr{U^\dagger \DCC{i} U \rho^{-1} V^\dagger \epsilon_i V \rho}.$$
 We will use an extended version of equations \eqref{eq:rhokappa}. Consider two diagonal matrices:
$$
G=\left(\begin{array}{ccc}
         g_1&0&0\\
	 0&g_2&0\\
	 0&0&g_3
        \end{array}
 \right),\quad H=\left(\begin{array}{ccc}
         h_1&0&0\\
	 0&h_2&0\\
	 0&0&h_3
        \end{array}
 \right).
$$
The following equations hold:
\begin{equation}\label{eq:GkH}
 G \kappa^+_1 H= g_1 h_2\, \kappa^+_1,\  G \kappa^-_1 H= g_2 h_1\, \kappa^-_1,\ G \kappa^+_2 H = g_3 h_1\, \kappa^+_2,\  G \kappa^-_2 H = g_1 h_3\, \kappa^-_2,\ G \kappa^+_3 H= g_2 h_3\, \kappa^+_3,\ G \kappa^-_3 H= g_3 h_2\, \kappa^-_3.
\end{equation}
We introduce six chiral fields, which are real and imaginary parts of the following vectors:
\begin{equation}\label{eq:uvw}
\uu=V \kappa^+_1 V^\dagger,\quad \vv=V \kappa^+_2 V^\dagger,\quad \ww=V \kappa^+_3 V^\dagger.
\end{equation}
Those fields, together with $\nn$ and $\mm$ are analogous to the triple of fields $\kb,\lm,\mb$ \eqref{eq:l},\eqref{eq:km} in the SU(2) case.
Using them, equations \eqref{eq:UdU} and \eqref{eq:VdV} we obtain:
\begin{eqnarray*}
\Tr{U^\dagger \DCC{i} U \rho \Proj( V^\dagger \partial_i V )\rho}=-\frac{1}{4}\rho_1 \rho_2 \left((\DCC{i} \n,[\kappa_1^+,\n])+(\DCC{i} \m,[\kappa_1^+,\m])-2 \sJ{i}{3}\, \n_1^+-2 \sJ{i}{8}\, \m_1^+\right)\left( (\partial_i\nn,[\uu^\dagger,\nn])+ (\partial_i\mm,[\uu^\dagger,\mm])\right)+\\+\frac{1}{4}\rho_1 \rho_2 \left((\DCC{i} \n,[\kappa_1^-,\n])+(\DCC{i} \m,[\kappa_1^-,\m])-2 \sJ{i}{3}\, \n_1^--2 \sJ{i}{8}\, \m_1^-\right)\left( (\partial_i\nn,[\uu,\nn])+ (\partial_i\mm,[\uu,\mm])\right)-\\
-\frac{1}{4}\rho_1 \rho_3 \left((\DCC{i} \n,[\kappa_2^+,\n])+(\DCC{i} \m,[\kappa_2^+,\m])-2 \sJ{i}{3}\, \n_2^+-2 \sJ{i}{8}\, \m_2^+\right)\left( (\partial_i\nn,[\vv^\dagger,\nn])+ (\partial_i\mm,[\vv^\dagger,\mm])\right)+\\+\frac{1}{4}\rho_1 \rho_3 \left((\DCC{i} \n,[\kappa_2^-,\n])+(\DCC{i} \m,[\kappa_2^-,\m])-2 \sJ{i}{3}\, \n_2^--2 \sJ{i}{8}\, \m_2^-\right)\left( (\partial_i\nn,[\vv,\nn])+ (\partial_i\mm,[\vv,\mm])\right)-\\
-\frac{1}{4}\rho_2 \rho_3 \left((\DCC{i} \n,[\kappa_3^+,\n])+(\DCC{i} \m,[\kappa_3^+,\m])-2 \sJ{i}{3}\, \n_3^+-2 \sJ{i}{8}\, \m_3^+\right)\left( (\partial_i\nn,[\ww^\dagger,\nn])+ (\partial_i\mm,[\ww^\dagger,\mm])\right)+\\+\frac{1}{4}\rho_2 \rho_3 \left((\DCC{i} \n,[\kappa_3^-,\n])+(\DCC{i} \m,[\kappa_3^-,\m])-2 \sJ{i}{3}\, \n_3^--2 \sJ{i}{8}\, \m_3^-\right)\left( (\partial_i\nn,[\ww,\nn])+ (\partial_i\mm,[\ww,\mm])\right)
,
\end{eqnarray*}
where $\n^\pm_i:=(\n,\kappa_i^\pm),\ \m^\pm_i:=(\m,\kappa_i^\pm)$.
Similarly, we calculate the remaining two terms:
\begin{eqnarray*}
 -\frac{\iu}{2} e^{3\iu \phi} \det (\rho)\Tr{U^\dagger \DCC{i} U \rho V^\dagger \epsilon_i V \rho^{-1}}+\frac{\iu}{2} e^{-3\iu \phi} \det (\rho)\Tr{U^\dagger \DCC{i} U \rho^{-1} V^\dagger \epsilon_i V \rho}=\\
=\frac{\iu}{4} \rho_3 (\rho_1^2 e^{3\iu \phi}-\rho_2^2 e^{-3\iu \phi}) \left((\DCC{i} \n,[\kappa_1^+,\n])+(\DCC{i} \m,[\kappa_1^+,\m])-2 \sJ{i}{3}\, \n_1^+-2 \sJ{i}{8}\, \m_1^+\right)(\epsilon_i, \uu^\dagger)-\\-\frac{\iu}{4} \rho_3(\rho_2^2 e^{3\iu \phi}-\rho_1^2 e^{-3\iu \phi})\left((\DCC{i} \n,[\kappa_1^-,\n])+(\DCC{i} \m,[\kappa_1^-,\m])-2 \sJ{i}{3}\, \n_1^--2 \sJ{i}{8}\, \m_1^-\right)(\epsilon_i, \uu)+\\
+\frac{\iu}{4} \rho_2(\rho_3^2 e^{3\iu \phi}-\rho_1^2 e^{-3\iu \phi})\left((\DCC{i} \n,[\kappa_2^+,\n])+(\DCC{i} \m,[\kappa_2^+,\m])-2 \sJ{i}{3}\, \n_2^+-2 \sJ{i}{8}\, \m_2^+\right) (\epsilon_i, \vv^\dagger)-\\-\frac{\iu}{4}\rho_2(\rho_1^2 e^{3\iu \phi}-\rho_3^2  e^{-3\iu \phi}) \left((\DCC{i} \n,[\kappa_2^-,\n])+(\DCC{i} \m,[\kappa_2^-,\m])-2 \sJ{i}{3}\, \n_2^--2 \sJ{i}{8}\, \m_2^-\right)(\epsilon_i, \vv)+\\+\frac{\iu}{4}\rho_1(\rho_2^2 e^{3\iu \phi}-\rho_3^2 e^{-3\iu \phi}) \left((\DCC{i} \n,[\kappa_3^+,\n])+(\DCC{i} \m,[\kappa_3^+,\m])-2 \sJ{i}{3}\, \n_3^+-2 \sJ{i}{8}\, \m_3^+\right) (\epsilon_i, \ww^\dagger)-\\-\frac{\iu}{4}\rho_1(\rho_3^2 e^{3\iu \phi}-\rho_2^2 e^{-3\iu \phi}) \left((\DCC{i} \n,[\kappa_3^-,\n])+(\DCC{i} \m,[\kappa_3^-,\m])-2 \sJ{i}{3}\, \n_3^--2 \sJ{i}{8}\, \m_3^-\right)(\epsilon_i, \ww)+\\+\frac{1}{2}\sin(3\phi)\det(\rho)\left((\DCC{i} \n,[\kappa_3,\n])+(\DCC{i} \m,[\kappa_3,\m])-2 \sJ{i}{3}\, \n_3-2 \sJ{i}{8}\, \m_3\right)(\epsilon_i,\cos(\alpha)\nn+\sin(\alpha)\mm)+\\+\frac{1}{2}\sin(
3\phi)\det(\rho)\left((\DCC{i} \n,[\kappa_8,\n])+(\DCC{i} \m,[\kappa_8,\m])-2 \sJ{i}{3}\, \n_8-2 \sJ{i}{8}\, \m_8\right)(\epsilon_i,-\sin(\alpha)\nn+\cos(\alpha)\mm),
\end{eqnarray*}
where $\n_3:=(\n,\kappa_3),\ \m_8:=(\m,\kappa_8)$.
\subsubsection{The term $-\iu \cos(3 \phi) \det(\rho) \Tr{\rho^{-1}\partial_i \rho V^\dagger \epsilon_i V}$}
In this subsection we study the term $-\iu \cos(3 \phi) \det(\rho) \Tr{\rho^{-1}\partial_i \rho V^\dagger \epsilon_i V}$ which is of zeroth order in derivatives of $\n,\,\m,\,\nn,\,\mm$. It modifies the dynamics of the scalar fields $\rho_1,\,\rho_2,\,\rho_3$. We decompose $-\iu \rho^{-1}\partial_i \rho$ in a basis of diagonal matrices: $\mathbbm{1},\ \kappa_3^\alpha,\ \kappa_8^\alpha$:
\begin{eqnarray*}
-\iu\rho^{-1}\partial_i \rho=-\frac{\iu}{3} \left(\frac{\partial_i \rho_1}{\rho_1}+\frac{\partial_i \rho_2}{\rho_2}+\frac{\partial_i \rho_3}{\rho_3}\right)\mathbbm{1}+\sqrt{\frac{2}{3}}\left(- \sin(\alpha-\frac{\pi}{3})\frac{\partial_i \rho_1}{\rho_1}-\sin(\alpha+\frac{\pi}{3})\frac{\partial_i \rho_2}{\rho_2} +\sin(\alpha) \frac{\partial_i \rho_3}{\rho_3} \right) \kappa_3^\alpha+\\+\sqrt{\frac{2}{3}}\left( \cos(\alpha-\frac{\pi}{3})\frac{\partial_i \rho_1}{\rho_1}+\cos(\alpha+\frac{\pi}{3})\frac{\partial_i \rho_2}{\rho_2} -\cos(\alpha) \frac{\partial_i \rho_3}{\rho_3} \right) \kappa_8^\alpha.
\end{eqnarray*}
As a result:
\begin{eqnarray*}
-\iu \cos(3 \phi) \det(\rho) \Tr{\rho^{-1}\partial_i \rho V^\dagger \epsilon_i V}=\sqrt{\frac{2}{3}}\cos(3 \phi) \det (\rho) \frac{\partial_i \rho_1}{\rho_1}\left(\sin(\alpha-\frac{\pi}{3})(\nn,\epsilon_i)- \cos(\alpha-\frac{\pi}{3}) (\mm,\epsilon_i)\right)+\\+\sqrt{\frac{2}{3}}\cos(3 \phi) \det (\rho) \frac{\partial_i \rho_2}{\rho_2}\left(\sin(\alpha+\frac{\pi}{3})(\nn,\epsilon_i)- \cos(\alpha+\frac{\pi}{3}) (\mm,\epsilon_i)\right)-\sqrt{\frac{2}{3}}\cos(3 \phi) \det (\rho) \frac{\partial_i \rho_3}{\rho_3}\left(\sin(\alpha)(\nn,\epsilon_i)- \cos(\alpha) (\mm,\epsilon_i)\right).
\end{eqnarray*}
Note, that there are only first order derivatives of the scalar fields. There are no analogous terms in the static SU(2) Lagrangian; in the SU(2) case there are only terms in second or in zeroth order of derivatives of the scalar field.
\subsection{The term $\frac{1}{4} \left(\partial_{[i} A^3_{j]}+2\Pt_{ij} \right)^2+\frac{1}{4} \left(\partial_{[i} A^8_{j]}+2\Pe_{ij} \right)^2-\frac{3}{4}(\Pt_{ij})^2-\frac{3}{4}(\Pe_{ij})^2$}
In this section we show, that the term $\frac{1}{4} \left(\partial_{[i} A^3_{j]}+2\Pt_{ij} \right)^2+\frac{1}{4} \left(\partial_{[i} A^8_{j]}+2\Pe_{ij} \right)^2$ gives rise to the Maxwell parts of the nonlinear sigma models. We also interpret the tensors $\Pt_{ij}$ and $\Pe_{ij}$.
\subsubsection{Maxwell terms}
As in the SU(2) case, we replace the gauge potentials $A^3_i$ and $A^8_i$ with the \UIUI\ and \UCUC\ gauge invariant supercurrents $\sJ{i}{3}$ and $\sJ{i}{8}$. The supercurrents are
$$
\sJ{i}{3}=\Tr{\kappa_3 U \partial_i U^\dagger} + \frac{3}{2} A^3_i + \frac{\sqrt{3}}{2} A^8_i + \Tr{\n \C_i},\quad \sJ{i}{8}=\Tr{\kappa_8 U \partial_i U^\dagger} + \frac{\sqrt{3}}{2} A^3_i - \frac{3}{2} A^8_i + \Tr{\m \C_i}.
$$
Therefore the gauge potentials $A^3_i$ and $A^8_i$ can be expressed in the following way:
\begin{eqnarray*}
A^3_i= \frac{1}{2} \left(\sJ{i}{3} -\Tr{\kappa_3 U \partial_i U^\dagger} -\Tr{\n \C_i}\right)+\frac{1}{2\sqrt{3}}\left(\sJ{i}{8} -\Tr{\kappa_8 U \partial_i U^\dagger} -\Tr{\m \C_i}\right),\\
A^8_i= \frac{1}{2\sqrt{3}} \left(\sJ{i}{3} -\Tr{\kappa_3 U \partial_i U^\dagger} -\Tr{\n \C_i}\right) -\frac{1}{2} \left(\sJ{i}{8} -\Tr{\kappa_8 U \partial_i U^\dagger} -\Tr{\m \C_i}\right).
\end{eqnarray*}
Next, we calculate the Maxwell tensors $\partial_{[i}A_{j]}^3$ and $\partial_{[i}A_{j]}^8$.

 First, notice that $\partial_{[i}\Tr{\kappa_3 U \partial_{j]} U^\dagger}$ and $\partial_{[i}\Tr{\kappa_8 U \partial_{j]} U^\dagger}$ are pullbacks of the Kirillov-Konstant symplectic forms on SU(3)/(U(1)$\times$U(1)) and on SU(3)/(SU(2)$\times$U(1)) respectively. They can be expressed using chiral fields $\n$ and $\m$ \cite{Faddeev_Niemi_SUN}:
$$
\partial_{[i} \Tr{\kappa_3 U \partial_{j]} U^\dagger}=\frac{1}{2}(\n,[\partial_i\n, \partial_j \n] +[\partial_i\m, \partial_j \m]),\quad \partial_{[i} \Tr{\kappa_8 U \partial_{j]} U^\dagger}=\frac{1}{2}(\m,[\partial_i\n, \partial_j \n] +[\partial_i\m, \partial_j \m]).
$$
Similarly, $\partial_{[i} \C_{j]}^3$ and $\partial_{[i} \C_{j]}^8$ are pullbacks of the Kirillov-Konstant symplectic forms on SU(3)/(U(1)$\times$U(1)) and SU(3)/(SU(2)$\times$U(1)) respectively (let us recall, that $\C_i^3=\Tr{\kappa_3 V^\dagger \partial_i V},\, \C_i^8=\Tr{\kappa_8 V^\dagger \partial_i V}$ \eqref{eq:C}). One can express them in terms of the chiral fields $\nn$ and $\mm$:
$$
\cos(\alpha)\,\partial_{[i} \C_{j]}^3-\sin(\alpha)\,\partial_{[i} \C_{j]}^8=\frac{1}{2}(\nn,[\partial_i\nn, \partial_j \nn] +[\partial_i\mm, \partial_j \mm]),\quad \sin(\alpha)\,\partial_{[i} \C_{j]}^3+\cos(\alpha)\,\partial_{[i} \C_{j]}^8=\frac{1}{2}(\mm,[\partial_i\nn, \partial_j \nn] +[\partial_i\mm, \partial_j \mm]).
$$
As a result
\begin{eqnarray*}
\partial_{[i} \C_{j]}^3=\frac{1}{2}\cos(\alpha)\,(\nn,[\partial_i\nn, \partial_j \nn] +[\partial_i\mm, \partial_j \mm])+\frac{1}{2}\sin(\alpha)\,(\mm,[\partial_i\nn, \partial_j \nn] +[\partial_i\mm, \partial_j \mm]),\\
\partial_{[i} \C_{j]}^8=-\frac{1}{2}\sin(\alpha)\,(\nn,[\partial_i\nn, \partial_j \nn] +[\partial_i\mm, \partial_j \mm])+\frac{1}{2}\cos(\alpha)\,(\mm,[\partial_i\nn, \partial_j \nn] +[\partial_i\mm, \partial_j \mm]).
\end{eqnarray*}
Taking into account, that
\begin{eqnarray*}
\frac{1}{2}(\n,[\DCC{i}\n, \DCC{j} \n] +[\DCC{i}\m, \DCC{j} \m])=\frac{1}{2}(\n,[\partial_i\n, \partial_j \n] +[\partial_i\m, \partial_j \m])-(\partial_{[i}\n, \C_{j]}),\\
\frac{1}{2}(\m,[\DCC{i}\n, \DCC{j} \n] +[\DCC{i}\m, \DCC{j} \m])=\frac{1}{2}(\m,[\partial_i\n, \partial_j \n] +[\partial_i\m, \partial_j \m])-(\partial_{[i}\m, \C_{j]}),
\end{eqnarray*}
we obtain:
\begin{eqnarray*}
 \partial_{[i}\left(\Tr{\kappa_3 U \partial_{j]} U^\dagger} +\Tr{\n \C_{j]}}\right)=\frac{1}{2}(\n,[\DCC{i}\n, \DCC{j} \n] +[\DCC{i}\m, \DCC{j} \m])+&\\+\frac{1}{2}\left(\cos(\alpha)\nna{3} -\sin(\alpha)\nna{8}\right)\,(\nn,[\partial_i\nn, \partial_j \nn] +[\partial_i\mm, \partial_j \mm])+\frac{1}{2}&\left(\sin(\alpha)\nna{3}+\cos(\alpha)\nna{8} \right)\,(\mm,[\partial_i\nn, \partial_j \nn] +[\partial_i\mm, \partial_j \mm]),\\
 \partial_{[i}\left(\Tr{\kappa_8 U \partial_{j]} U^\dagger} +\Tr{\m \C_{j]}}\right)=\frac{1}{2}(\m,[\DCC{i}\n, \DCC{j} \n] +[\DCC{i}\m, \DCC{j} \m])+&\\+\frac{1}{2}\left(\cos(\alpha)\mma{3} -\sin(\alpha)\mma{8}\right)\,(\nn,[\partial_i\nn, \partial_j \nn] +[\partial_i\mm, \partial_j \mm])+\frac{1}{2}&\left(\sin(\alpha)\mma{3}+\cos(\alpha)\mma{8} \right)\,(\mm,[\partial_i\nn, \partial_j \nn] +[\partial_i\mm, \partial_j \mm])
\end{eqnarray*}
Finally, the expression for the terms investigated in this subsection is:
\begin{eqnarray}
&\frac{1}{4} \left(\partial_{[i} A^3_{j]}+2\Pt_{ij} \right)^2+\frac{1}{4} \left(\partial_{[i} A^8_{j]}+2\Pe_{ij} \right)^2-\frac{3}{4}(\Pt_{ij})^2-\frac{3}{4}(\Pe_{ij})^2=\label{eq:Maxwell_SUt}\\&=\frac{1}{12}\, (\Mt_{ij}-\n^\alpha_3\,\Kt_{ij}-\n^\alpha_8\,\Ke_{ij} + \Lt_{ij} - 6\Ptt_{ij})^2+\frac{1}{12}\, (\Me_{ij}-\m^\alpha_3\,\Kt_{ij}-\m^\alpha_8\,\Ke_{ij} + \Le_{ij} - 6\Pee_{ij})^2-9\, (\Ptt_{ij})^2 - 9\, (\Pee_{ij})^2,\nonumber
\end{eqnarray}
where 
\begin{eqnarray*}&\Mt_{ij}:=\frac{1}{2}(\n,[\DCC{i}\n, \DCC{j} \n] +[\DCC{i}\m, \DCC{j} \m]),\quad \Me_{ij}:=\frac{1}{2}(\m,[\DCC{i}\n, \DCC{j} \n] +[\DCC{i}\m, \DCC{j} \m]),\\& \Kt_{ij}:=-\frac{1}{2}(\nn,[\partial_i\nn, \partial_j \nn]+[\partial_i\mm, \partial_j \mm]),\quad \Ke_{ij}:=-\frac{1}{2}(\mm,[\partial_i\nn, \partial_j \nn]+[\partial_i\mm, \partial_j \mm]),\quad \Lt_{ij}:=-\partial_{[i}\sJ{{j]}}{3},\quad \Le_{ij}:=-\partial_{[i}\sJ{{j]}}{8},\\&
 \n^\alpha_3:=(\n,\kappa^\alpha_3),\ \n^\alpha_8:=(\n,\kappa^\alpha_8),\ \m^\alpha_3:=(\m,\kappa^\alpha_3),\ \m^\alpha_8:=(\m,\kappa^\alpha_8),\ \Ptt_{ij}:=\frac{1}{4\sqrt{3}}(\sqrt{3}\Pt_{ij}+\Pe_{ij}),\ \Pee_{ij}:=\frac{1}{4\sqrt{3}}(\Pt_{ij}-\sqrt{3}\Pe_{ij}).
\end{eqnarray*}
 The formula should be compared with the first term in the expression \eqref{eq:sigma} for the Lagrangian of SU(2) Yang-Mills theory. This equation reveals \textbf{Maxwell terms} $(\Mt_{ij})^2+(\Me_{ij})^2$ and $(\Kt_{ij})^2+(\Ke_{ij})^2$ of the nonlinear sigma models. As a result this formula together with the equations \eqref{eq:sigma1} and \eqref{eq:sigma2} describe how the expected sigma model \eqref{eq:action} appears in the static SU(3) Yang-Mills Lagrangian. It appears twice, once for $\n$ and $\m$ fields, and second time for $\nn$ and $\mm$ fields. It is easy to notice a duality between the $\n$, $\m$ fields and the $\nn$, $\mm$ fields. It is analogous to the duality between the $\tttt$ and $\lv$ fields in the SU(2) case.
\subsubsection{The tensors $\Pt$ and $\Pe$}
The tensors $\Pt_{ij}$ and $\Pe_{ij}$ are analogous to the tensor $P_{ij}$ from the SU(2) case ($P_{ij}$ is studied in section \ref{sc:P}). Similarly to the SU(2) case, we interpret the terms using the chiral fields and the scalar fields.

As in the SU(2) case, it will be convenient for us to use the following tensors:
$$
\Pt_k:=\epsilon_{ijk} \Pt_{ij},\quad \Pe_k:=\epsilon_{ijk} \Pe_{ij}.
$$
The relations can be inverted:
$$
\Pt_{ij}=\frac{1}{2}\epsilon_{ijk} \Pt_k,\quad \Pe_{ij}=\frac{1}{2}\epsilon_{ijk} \Pe_k.
$$
We use matrix interpretation of $B$, and obtain, that:
$$
\Pt_{k}=\Tr{\tau_1 B \epsilon_k B^\dagger},\quad \Pe_{k}=\Tr{\tau_2 B \epsilon_k B^\dagger},
$$
where
$$
\tau_1=\left[ \begin{array}{ccc} -i2 \sqrt{2}&0&0\\ 0&i\sqrt{2}&0\\0&0&i\sqrt{2} \end{array}\right],\quad \tau_2=\left[ \begin{array}{ccc} 0&0&0\\ 0&i\sqrt{6}&0\\0&0&-i\sqrt{6} \end{array}\right].
$$
The matrices $\tau_1$ and $\tau_2$ can be written in the basis $\kappa_3,\, \kappa_8$:
$$
\tau_1=3\,\kappa_3 + \sqrt{3}\, \kappa_8,\quad \tau_2=\sqrt{3}\,\kappa_3 - 3\, \kappa_8.
$$
As a result, 
\begin{equation}
\Pt_k=\Tr{U^\dagger \left(3\,\kappa_3 + \sqrt{3}\, \kappa_8\right) U  \rho V^\dagger \epsilon_{k} V \rho},\quad \Pe_k= \Tr{U^\dagger \left(\sqrt{3}\,\kappa_3 - 3\, \kappa_8\right) U  \rho V^\dagger \epsilon_{k} V \rho}.
\end{equation}
Let us recall, that $\Ptt_{ij}:=\frac{1}{4\sqrt{3}}(\sqrt{3}\Pt_{ij}+\Pe_{ij}),\ \Pee_{ij}:=\frac{1}{4\sqrt{3}}(\Pt_{ij}-\sqrt{3}\Pe_{ij})$. 
We define:
$$
\Ptt_{k}:= \epsilon_{ijk}\Ptt_{ij}, \quad \Pee_{k}:= \epsilon_{ijk}\Pee_{ij}.
$$
Explicitly: 
\begin{equation}
 \Ptt_k=\Tr{\n  \rho V^\dagger \epsilon_{k} V \rho},\quad \Pee_k=\Tr{\m \rho V^\dagger \epsilon_{k} V \rho}.
\end{equation}
Using the equation \eqref{eq:GkH} the tensors $\Ptt_{ij}$ and $\Pee_{ij}$ can be expressed in terms of fields $\rho,\n,\m,\nn,\mm,\uu,\,\vv,\,\ww$ (let us recall, that the latter three fields are defined in \eqref{eq:uvw}):
\begin{eqnarray*}
&\Ptt_{k}= - \rho_1 \rho_2 (\n^+_1\uu_k- \n^-_1\uu^\dagger_k) - \rho_1 \rho_3 ( \n^+_2 \vv_k - \n^-_2 \vv^\dagger_k) - \rho_2 \rho_3 ( \n^+_3 \ww_k - \n^-_3 \ww^\dagger_k)+\\&+\frac{1}{\sqrt{3}} \left(\rho_1^2\sin(\alpha-\frac{\pi}{3})-\rho_2^2 \sin(\alpha+\frac{\pi}{3}) \right)\n_3 \nn_k - \frac{1}{\sqrt{3}} \left( \rho_1^2 \cos(\alpha-\frac{\pi}{3})-\rho_2^2 \cos(\alpha+\frac{\pi}{3}) \right) \n_3 \mm_k+\\&+\frac{1}{3}\left(\rho_1^2 \sin(\alpha-\frac{\pi}{3}) +\rho_2^2 \sin(\alpha+\frac{\pi}{3}) +2 \rho_3^2 \sin(\alpha) \right) \n_8 \nn_k -\frac{1}{3}\left(\rho_1^2 \cos(\alpha-\frac{\pi}{3}) + \rho_2^2 \cos(\alpha+\frac{\pi}{3})+2\rho_3^2 \cos(\alpha)\right)\n_8\mm_k,\\
&\Pee_{k}= - \rho_1 \rho_2 (\m^+_1\uu_k- \m^-_1\uu^\dagger_k) - \rho_1 \rho_3 ( \m^+_2 \vv_k - \m^-_2 \vv^\dagger_k) - \rho_2 \rho_3 ( \m^+_3 \ww_k - \m^-_3 \ww^\dagger_k)+\\&+\frac{1}{\sqrt{3}} \left(\rho_1^2\sin(\alpha-\frac{\pi}{3})-\rho_2^2 \sin(\alpha+\frac{\pi}{3}) \right)\m_3 \nn_k - \frac{1}{\sqrt{3}} \left( \rho_1^2 \cos(\alpha-\frac{\pi}{3})-\rho_2^2 \cos(\alpha+\frac{\pi}{3}) \right) \m_3 \mm_k+\\&+\frac{1}{3}\left(\rho_1^2 \sin(\alpha-\frac{\pi}{3}) +\rho_2^2 \sin(\alpha+\frac{\pi}{3}) +2 \rho_3^2 \sin(\alpha) \right) \m_8 \nn_k -\frac{1}{3}\left(\rho_1^2 \cos(\alpha-\frac{\pi}{3}) + \rho_2^2 \cos(\alpha+\frac{\pi}{3})+2\rho_3^2 \cos(\alpha)\right)\m_8\mm_k,\\
\end{eqnarray*}
where $\n^\pm_p=(\n,\kappa^\pm_p),\m^\pm_p=(\m,\kappa^\pm_p),\, p\in\{1,2,3\},\n_3=(\n,\kappa_3),\,\n_8=(\n,\kappa_8),\,\m_3=(\n,\kappa_3),\, \m_8=(m,\kappa_8)$, $\uu_k=(\uu,\epsilon_k),\, \vv_k=(\vv,\epsilon_k),\, \ww_k=(\ww,\epsilon_k)$, $\, \nn_k=(\nn,\epsilon_k),\, \mm_k=(\mm,\epsilon_k)$.

Note, that the expression above includes not only the chiral fields $\n,\,\m,\,\nn,\,\mm$ and the scalar fields $\rho_1,\,\rho_2,\,\rho_3$, but also the chiral fields $\uu,\,\vv,\,\ww$. It is in contrast with the SU(2) case, where the fields $\kb,\,\mb$ (the fields $\uu,\,\vv,\,\ww$ are analogous to the fields $\kb,\,\mb$) were not present in the expression for $P_{ij}$. It is caused by different matrix structures of the decompositions: $U$ and $V$ are $3\times 3$ matrices, whereas $\Q$ is a co-vector and $\SO$ is a $3\times 3$ matrix.
\section{Summary}{\label{sc:summary}}
We introduced new variables for static SU(3) Yang-Mills theory. The idea is to treat the field $B$ -- the part of the gauge potential orthogonal to the Cartan subalgebra of su(3) -- as a complex 3$\times$3 matrix. Then singular value decomposition can be applied, and the decomposition is:
\begin{equation}\label{eq:BUrV}B= U \rho V^\dagger,\end{equation}
where $U(x)$ is a U(3) matrix, $V(x)$ is an SU(3) matrices and $\rho(x)$ is diagonal $3\times 3$ matrix. In those variables, the static Lagrangian takes the following form ($A^3_i$ and $A^8_i$ are the components in the direction of the Cartan subalgebra):
\begin{eqnarray}
 &L_{SU(3)}=-\frac{1}{2}\Tr{U^\dagger \DCC{i} U \rho^2  U^\dagger \DCC{i} U}- \frac{1}{2} \Tr{\Proj(V^\dagger \partial_i V) \rho^2 \Proj(V^\dagger \partial_i V)}+\Tr{U^\dagger \DCC{i} U \rho \Proj( V^\dagger \partial_i V )\rho}+\nonumber\\&+\frac{1}{4} \left(\partial_{[i} A^3_{j]}+2\Pt_{ij} \right)^2+\frac{1}{4} \left(\partial_{[i} A^8_{j]}+2\Pe_{ij} \right)^2+\textrm{terms depending on global phase shifting}+\nonumber\\&+\frac{1}{2} \Tr{\partial_i \rho \partial_i \rho} + \det (\rho)^2 \Tr{\rho^{-2}}- \Tr{\rho^2}^2+ \Tr{\rho^4}-\frac{3}{4}(\Pt_{ij})^2-\frac{3}{4}(\Pe_{ij})^2.\label{eq:LSUt}
\end{eqnarray}
The terms depending on the global phase shifting of the field $B_i^p$ come from the fact, that the term $D_{i} B_{j}^p$ and the term $\iu \epsilon^{pqr} \overline{B^q_i B^r_j}$ in the static Lagrangian \eqref{eq:lagrangian} transform differently under the global phase shifting. There is no analogous term to $\iu \epsilon^{pqr} \overline{B^q_i B^r_j}$ in SU(2) Yang-Mills theory, and these terms can not be compared to any terms in the SU(2) case. The terms depending on global phase shifting are:
\begin{itemize}
 \item interaction term that is of first order in derivatives of the chiral fields $\n$ and $\m$:
$$
-\frac{\iu}{2} e^{3\iu \phi} \det (\rho)\Tr{U^\dagger \DCC{i} U \rho V^\dagger \epsilon_i V \rho^{-1}}+\frac{\iu}{2} e^{-3\iu \phi} \det (\rho)\Tr{U^\dagger \DCC{i} U \rho^{-1} V^\dagger \epsilon_i V \rho};
$$
\item term in first order of derivatives of the chiral fields $\nn$ and $\mm$:
$$
-\sin(3\phi) \det(\rho) \Tr{\Proj(V^\dagger\partial_i V) V^\dagger \epsilon_i V};
$$
\item term in first order of derivatives of the field $\rho$:
$$
-\iu \cos(3 \phi) \det(\rho) \Tr{\rho^{-1}\partial_i \rho V^\dagger \epsilon_i V}.
$$
\end{itemize}

The analysis of the static SU(3) Lagrangian is based on the observation, that the Faddeev-Niemi decomposition of the SU(2) connection can be written in a similar way:
\begin{equation}\label{eq:XQrO}
X^+ = \Q\, \rho\, \SO,
\end{equation}
where the complex co-vector $X^+(x)$ is the part of the connection orthogonal to the Cartan subalgebra, $\Q(x)$ is normalized co-vector, and therefore defines an SU(2) matrix $Q(x)$, $\rho(x)$ is a scalar, and $\SO(x)$ is an SO(3) matrix. In those variables the static SU(2) Yang-Mills Lagrangian takes the following form:
\begin{equation}\label{eq:LSUd}
L_{SU(2)}=-\frac{\rho^2}{4} \Tr{Q\DC{i}Q^\dagger Q\DC{i}Q^\dagger}-\frac{\rho^2}{2}\Tr{(\Q^\dagger\otimes \Q) \proj( \SO \partial_i \SO^T) \proj( \SO \partial_i \SO^T)}+ \frac{1}{4}( \partial_{[i} A_{j]} + 2 P_{ij})^2+ \frac{1}{2} (\partial_i \rho)^2  -\frac{3}{4} P_{ij}^2.
\end{equation}
 Thanks to the structural similarity of the decompositions \eqref{eq:BUrV} and \eqref{eq:XQrO} as well as the structural similarity of the static Lagrangians \eqref{eq:LSUt} and \eqref{eq:LSUd} many properties of the decomposition we introduced are analogous to the properties of the Faddeev-Niemi decomposition. We briefly summarize them in the table below (table \ref{tab:podsumowanie}). Importantly, note that there is a duality between the fields $\n$, $\m$ and the fields $\nn$, $\mm$. This is a property analogous to the duality between the $\ttt$ and $\lv$ fields in the SU(2) case.  For both pairs of fields  $\n$, $\m$ and $\nn$, $\mm$ there are terms in the Lagrangian that may support knot-like excitations (in a similar way the nonlinear sigma model terms appear for the fields $\ttt$ and $\lv$ in the SU(2) case). The chiral model relevant for the SU(3) Yang-Mills theory turns out to be the model anticipated by Faddeev and Niemi and studied by Shabanov \cite{Faddeev_Niemi_SUN,
Shabanov} (defined by the action \eqref{eq:action}).

Note that there is no term in the SU(2) case analogous to the term $\Tr{U^\dagger \DCC{i} U \rho \Proj( V^\dagger \partial_i V )\rho}$ in the SU(3) case. It is due to the matrix structure of the decompositions: in the decomposition \eqref{eq:BUrV} $U(x)$ and $V(x)$ are $3\times 3$ matrices, whereas in the decomposition \eqref{eq:XQrO} $\Q(x)$ is a co-vector and $\SO(x)$ is a $3\times 3$ matrix (in order to understand this phenomenon in details, see calculations in section \ref{sc:QQDODO}). As a result in the SU(3) case there are terms that describe interaction of the two chiral models and are in second order of derivatives of the chiral fields, first order in derivatives of the chiral fields $\n,\,\m$ and first order in derivatives of the chiral fields $\nn,\,\mm$. There is a duality between $\n$, $\m$ and $\nn$, $\mm$ in these terms. Let us note that in the SU(2) case terms involving (at least first order) derivatives of both chiral fields $\ttt$ and $\lv$ appear only in the Maxwell part (i.e. in the first 
term of 
expression \eqref{eq:sigma}). In the SU(3) case, analogous terms appear in the Maxwell term \eqref{eq:lagrangian}; additionally the term $\Tr{U^\dagger \DCC{i} U \rho \Proj( V^\dagger \partial_i V )\rho}$ describes interaction of the two chiral models in the part of the standard sigma model Hamiltonian.

\begin{table}[ht!]
 \centering
\begin{tabular}{|m{0.15\textwidth}|>{\centering}m{0.4\textwidth}<{\centering}|>{\centering}m{0.4\textwidth}<{\centering}|}
\hline
Property&SU(2) Yang-Mills theory& SU(3) Yang-Mills theory\tabularnewline
\hline
Decomposition of the connection&$X^+ = \Q\, \rho\, \SO$&$B=U \rho V^\dagger$\tabularnewline
\hline
External gauge symmetry& \UC & \UCUC\tabularnewline
\hline
Internal gauge symmetry& \UI & \UIUI\tabularnewline
\hline
Gauge potential(s) corresponding to internal gauge symmetry & $C_i=\frac{1}{2} \Tr{\epsilon_3 \SO \partial_i \SO^T}$ & $ \C_i^3=\Tr{\kappa_3 V^\dagger \partial_i V},\quad  \C_i^8=\Tr{\kappa_8 V^\dagger \partial_i V}$ \tabularnewline
\hline
Chiral fields& $$\tttt:=\frac{1}{\sqrt{2}}Q \sigma_3 Q^\dagger,$$ $$\lm= \frac{1}{\sqrt{2}} \SO^T \epsilon_3 \SO$$& $$\n=U^\dagger \kappa_3 U, \quad \m=U^\dagger \kappa_8 U,$$ $$\nn:= V\kappa_3^\alpha V^\dagger,\quad \mm:=V\kappa_8^\alpha V^\dagger$$\tabularnewline
\hline
Supercurrent(s)&$$J_i=\frac{\iu}{2} \Tr{\sigma_3 Q^\dagger \DC{i} Q}$$& $$\sJ{i}{3}= \Tr{\kappa_3 U \DCC{i} U^\dagger}, \quad \sJ{i}{8}= \Tr{\kappa_8 U \DCC{i} U^\dagger}$$
\tabularnewline
\hline
Standard sigma model Hamiltonian and mass terms for supercurrents& $$
-\frac{\rho^2}{4} \Tr{Q\DC{i}Q^\dagger Q \DC{i}Q^\dagger} =\frac{\rho^2}{8} (\DC{i} \tttt)^2 + \frac{\rho^2}{2} (J_i)^2
$$ & \begin{eqnarray*}
-\frac{1}{2}\Tr{U^\dagger \DCC{i} U \rho^2  U^\dagger \DCC{i}U}=\frac{1}{2}\chi^2 \left( \DCC{i} \n, \DCC{i}\n \right)+\\+\frac{1}{2}\chi^2 \left( \DCC{i} \m, \DCC{i}\m \right)+\chi^2(\sJ{i}{3})^2+\chi^2(\sJ{i}{8})^2+\\
+3\chi^2 (\partial_i \phi)^2+\textrm{terms not depending on $\chi^2$}
\end{eqnarray*}\tabularnewline\cline{2-3}
&\begin{eqnarray*}
&-\frac{\rho^2}{2}\Tr{(\Q^\dagger\otimes \Q) \proj(\SO \partial_i \SO^T)\proj( \SO \partial_i \SO^T)}=\\&=\frac{\rho^2}{4} (\partial_i \lv)^2+\frac{\rho^2}{4}\left((\Tr{\kb\partial_i \lm})^2-(\Tr{\mb\partial_i \lm})^2 \right)\tttt_3+\\&+\frac{\rho^2}{2} \Tr{\kb\partial_i\lm}\Tr{\mb\partial_i\lm} \tttt_1
\end{eqnarray*}& \begin{eqnarray*}
&-\frac{1}{2}\Tr{\Proj(V^\dagger \partial_i V) \rho^2 \Proj(V^\dagger \partial_i V)}=\\&=\frac{1}{2}\chi^2\,\left(\partial_i \nn,\partial_i \nn\right)+\frac{1}{2}\chi^2\,\left(\partial_i \mm,\partial_i \mm\right)+\\&+\frac{1}{2}r\,\left(\partial_i \nn,\partial_i \nn\right)-\frac{1}{2}r\,\left(\partial_i \mm,\partial_i \mm\right)-\chi^2 (\partial_i \alpha)^2
\end{eqnarray*}\tabularnewline
\hline
Maxwell terms&$$\frac{1}{4}( F_{ij} + 2 P_{ij})^2 =\frac{1}{4}(L_{ij}+M_{ij}-\tttt_2 K_{ij}-2 P_{ij})^2,$$
where
\begin{eqnarray*}&M_{ij}=\frac{1}{2}\tttt \cdot (\DC{i}\tttt \times \DC{j} \tttt),\\& K_{ij}=-\lv\cdot (\partial_i \lv \times \partial_j\lv),\quad L_{ij}=\partial_{[i} J_{j]}\end{eqnarray*}
& \begin{eqnarray*}
\frac{1}{4} \left(\partial_{[i} A^3_{j]}+2\Pt_{ij} \right)^2+\frac{1}{4} \left(\partial_{[i} A^8_{j]}+2\Pe_{ij} \right)^2=\\=\frac{1}{12}\, (\Lt_{ij}+\Mt_{ij}-\n^\alpha_3\,\Kt_{ij}-\n^\alpha_8\,\Ke_{ij} - 6\Ptt_{ij})^2+\\+\frac{1}{12}\, (\Le_{ij}+\Me_{ij}-\m^\alpha_3\,\Kt_{ij}-\m^\alpha_8\,\Ke_{ij} - 6\Pee_{ij})^2,
\end{eqnarray*}
where
\begin{eqnarray*}&\Mt_{ij}:=\frac{1}{2}(\n,[\DCC{i}\n, \DCC{j} \n] +[\DCC{i}\m, \DCC{j} \m]),\\& \Me_{ij}:=\frac{1}{2}(\m,[\DCC{i}\n, \DCC{j} \n] +[\DCC{i}\m, \DCC{j} \m]),\\& \Kt_{ij}:=-\frac{1}{2}(\nn,[\partial_i\nn, \partial_j \nn]+[\partial_i\mm, \partial_j \mm]),\  \Lt_{ij}:=-\partial_{[i}\sJ{{j]}}{3}\\& \Ke_{ij}:=-\frac{1}{2}(\mm,[\partial_i\nn, \partial_j \nn]+[\partial_i\mm, \partial_j \mm]),\   \Le_{ij}:=-\partial_{[i}\sJ{{j]}}{8}.
\end{eqnarray*}
\tabularnewline
\hline
\end{tabular}
\caption{Summary of the main properties of the decompositions.}
\label{tab:podsumowanie}
\end{table}
\section{Discussion and Outlook}\label{sc:discussion}
In our decomposition the chiral fields take values in SU(3)/(SU(2)$\times$U(1)) and SU(3)/(U(1)$\times$U(1)). Thanks to the fact, that $\pi_3$(SU(3)/(U(1)$\times$U(1)))=$\mathbb{Z}$, the variables may support knot-like excitations. This property is shared with the variables introduced by Bolokhov and Faddeev \cite{Bolokhov_Faddeev}, where the chiral fields take values in $\mathbb{S}^2$. Other decompositions were proposed by Kondo et al. \cite{Kondo} and  Evslin et al. \cite{Evslin}, however there the chiral field takes values in SU(3)/(SU(2)$\times$U(1)), and therefore can not support the knot-like scenario.

In our approach three scalar fields appear; they are eigenvalues of the 3x3 matrix $B^p_{\mu} \overline{B}^q_{\nu} \eta^{\mu\nu}$. The Maximal Abelian Gauge fixing (MAG) condition is the one, which describes gauge orbit extrema of $\int d^4x\, \Tr{B^{\mu} \overline{B}_{\mu}}$. Therefore the extrema of $\int d^4 x\, \Tr{B^{\mu} \overline{B}_{\mu}}$ are gauge invariant, and the field $\Tr{B^{\mu} \overline{B}_{\mu}}$ has an interpretation of a condensate. When the MAG condition is applied, the eigenvalues of $B^p_\mu \overline{B}^q_\nu \eta^{\mu\nu}$ are already U(1)$\times$U(1) gauge invariant. It would be therefore interesting to investigate the possibility of an interpretation of the three scalar fields as condensates. In such interpretation, the model relevant for infrared limit of SU(3) Yang-Mills theory may be an extension of the Faddeev-Niemi-Shabanov model (see \eqref{eq:nnmm}, \eqref{eq:sigmaext}, and a discussion after \eqref{eq:sigmaext}): the extension 
allows a linear combination of the standard 
sigma model Hamiltonian terms (with in general different positive coefficients) \eqref{eq:sigmaext}, and a SO(2) rotation of the Cartan subalgebra of su(3) \eqref{eq:nnmm}.

It would be interesting to extend the decomposition to (static) SU(N) Yang-Mills theory. In this case the matrix $B(x)$ is a $d\times 3$ matrix, where $d=\frac{N(N-1)}{2}$ is the number of the root vectors. The matrix $U(x)$ is a ${\rm U}(d)$ matrix, the matrix $\rho(x)$ is a $d\times 3$ rectangular diagonal matrix, and $V(x)$ is an SU(3) matrix. The external gauge symmetry is $({\rm U}_{\rm C}(1))^{N-1}$ and the internal gauge symmetry is: ${\rm U}_{\rm I}(1)\times{\rm U}_{\rm I}(1)\times {\rm U}_{\rm I}(d-3)$ when $N\geq 3$, \UI\ when $N=2$. The ${\rm U}_{\rm I}(d-3)$ part of the symmetry leaves $V(x)$ intact and transforms non-trivially only $U(x)$. In the case $N\geq 3$ the number of the degrees of freedom is $d^2+3+8-(d-3)^2- 2=6d$, and is equal to the number of degrees of freedom of the complex $d\times 3$ matrix $B(x)$.

In this article we considered only the static case. It would be interesting to extend the ideas to non-static case. One could try to extend further the analogy with the Faddeev-Niemi decomposition for the SU(2) Yang-Mills theory. There is however, yet another possible avenue to pursue, suggested by the considerations in section \ref{sec:interpretation}: one could consider the decomposition of the form $ B=U\, \rho\, V^\dagger,$ where $U(x)\in{\rm U}(3)$, $\rho(x)$ is a $3\times 4$ rectangular diagonal matrix, $V(x)\in\,$SU(4) in Euclidean case or $V(x)\in\,$SU(1,3) in Lorentzian case. There would be \UCUC\ external gauge symmetry and ${\rm U}_{\rm I}(1)\times{\rm U}_{\rm I}(1)\times {\rm U}_{\rm I}(1)$ internal gauge symmetry. The number of degrees of freedom of the decomposition is therefore $9+3+15-3=24$, and is equal to the number of degrees of freedom of the complex $3\times 4$ matrix $B(x)$.
\section*{Acknowledgments}
I gratefully acknowledge discussions with L. D. Faddeev. I would like to also thank for the hospitality I received at St. Petersburg Department of Steklov Mathematical Institute. This work was supported by the Foundation for Polish Science International PhD Projects Programme co-financed by the EU European Regional Development Fund.
\appendix
\section{The SU(3) Yang-Mills Lagrangian in Cartan-Weyl basis}\label{sc:YMCW}
In the Cartan-Weyl basis the field strength tensor takes the following form:
\begin{eqnarray*}
 \F^{+ 1}= dB^1-i\sqrt{2}A^3\wedge B^1+i \overline{B^2}\wedge \overline{B^3},& \F^{- 1}= d\overline{B^1}+i\sqrt{2}A^3\wedge \overline{B^1}-i {B^2}\wedge {B^3}\\
 \F^{+ 2}= dB^2+i\frac{\sqrt{2}}{2}A^3\wedge B^2+i\frac{\sqrt{6}}{2}A^8\wedge B^2+i \overline{B^3}\wedge \overline{B^1},&\quad \F^{- 2}= d\overline{B^2}-i\frac{\sqrt{2}}{2}A^3\wedge \overline{B^2}-i\frac{\sqrt{6}}{2}A^8\wedge \overline{B^2}-i {B^3}\wedge {B^1}\\
 \F^{+ 3}= dB^3+i\frac{\sqrt{2}}{2}A^3\wedge B^3-i\frac{\sqrt{6}}{2}A^8\wedge B^3+i \overline{B^1}\wedge \overline{B^2},&\quad \F^{- 3}= d\overline{B^3}-i\frac{\sqrt{2}}{2}A^3\wedge \overline{B^3}+i\frac{\sqrt{6}}{2}A^8\wedge \overline{B^3}-i {B^1}\wedge {B^2}\\
 \F^{3}=dA^3-i\sqrt{2} B^1\wedge \overline{B^1}+i\frac{\sqrt{2}}{2} B^2\wedge \overline{B^2}+i\frac{\sqrt{2}}{2} B^3\wedge \overline{B^3},&\quad F^8=dA^8+i\frac{\sqrt{6}}{2} B^2\wedge \overline{B^2}-i\frac{\sqrt{6}}{2} B^3\wedge \overline{B^3}
\end{eqnarray*}

We will partially fix the gauge -- we will use Maximal Abelian gauge fixing. Therefore, we introduce covariant derivative $D$ with respect to the remaining U(1)$\times$U(1) gauge transformations:
\begin{eqnarray*}
 D_\mu B^1_\nu:=\partial_\mu B^1_\nu-i\sqrt{2}A^3_\mu B^1_\nu,\ D_{\mu} B^2_\nu:=\partial_\mu B^2_\nu+i\frac{\sqrt{2}}{2}A^3_\mu B^2_\nu+i\frac{\sqrt{6}}{2}A^8_\mu B^2_\nu,\ D_\mu B^3_\nu:=\partial_\mu B^3_\nu+i\frac{\sqrt{2}}{2}A^3_\mu B^3_\nu-i\frac{\sqrt{6}}{2}A^8_\mu B^3_\nu,\\
 D_\mu \overline{B^1_\nu}:=\partial_\mu\overline{B^1_\nu}+i\sqrt{2}A^3_\mu \overline{B^1_\nu},\ D_\mu \overline{B^2_\nu}:=\partial_\mu\overline{B^2_\nu}-i\frac{\sqrt{2}}{2}A^3_\mu \overline{B^2_\nu}-i\frac{\sqrt{6}}{2}A^8_\mu \overline{B^2_\nu},\ D_\mu\overline{B^3_\nu}:=\partial_\mu\overline{B^3_\nu}-i\frac{\sqrt{2}}{2}A^3_\mu \overline{B^3_\nu}+i\frac{\sqrt{6}}{2}A^8_\mu \overline{B^3_\nu}.
\end{eqnarray*}

The Lagrangian of Yang-Mills theory with gauge fixing terms (and ghost part omitted) is:
\begin{equation*}
 L=\frac{1}{4} \F^{+ p}_{\mu \nu} \F^{- p\,\mu \nu}+\frac{1}{4} \F^{3}_{\mu \nu} \F^{3\,\mu \nu}+\frac{1}{4} \F^{8}_{\mu \nu} \F^{8\,\mu \nu}+\frac{\xi}{2}D^{\mu} B^p_\mu D^{\nu} \overline{B^p_\nu}.
\end{equation*}
In other words
\begin{equation*}
 L=\frac{1}{4} \left(D_{[\mu} B_{\nu]}^p+ i \epsilon^{pqr} \overline{B^q_{\mu} B^r_{\nu}} \right)\left( D^{[\mu} \overline{B^{p\,\nu]}}-i\epsilon^{pqr} B^q_{\mu} B^r_{\nu}\right)+\frac{1}{4} \F^{3}_{\mu \nu} \F^{3\,\mu \nu}+\frac{1}{4} \F^{8}_{\mu \nu} \F^{8\,\mu \nu}+\frac{\xi}{2}D^{\mu} B^p_\mu D^{\nu} \overline{B^p_\nu},
\end{equation*}
where $A_{[\mu\nu]}:=A_{\mu\nu}-A_{\nu\mu}$ is the antisymmetrization of indices.

The Lagrangian may be rewritten in the following way:
\begin{eqnarray*}
& L=\frac{1}{2} \left(D_{\mu} B_{\nu}^p +i \epsilon^{pqr} \overline{B^q_{\mu} B^r_{\nu}} \right)\left( D^{\mu} \overline{B^{p\,\nu}}-i\epsilon^{pqr} B^q_{\mu} B^r_{\nu}\right)- (B^q_\mu \overline{B^q_\mu})^2+ B^r_\mu \overline{B^q_\mu} B^q_\nu \overline{B^r_\nu}+\\& +\frac{1}{4} \F^{3}_{\mu \nu} \F^{3\,\mu \nu}+\frac{1}{4} \F^{8}_{\mu \nu} \F^{8\,\mu \nu}+\frac{\xi}{2}D^{\mu} B^p_\mu D^{\nu} \overline{B^p_\nu} - \frac{1}{2} D_{\mu} B^p_\nu D^{\nu} \overline{B^{p\,\mu}}.
\end{eqnarray*}
As pointed out in \cite{Bolokhov_Faddeev}, for $\xi=1$ the terms $\frac{\xi}{2}D^{\mu} B^p_\mu D^{\nu} \overline{B^p_\nu} - \frac{1}{2} D_{\mu} B^p_\nu D^{\nu} \overline{B^{p\,\mu}}$ double the cross terms in $\frac{1}{4} F^{3}_{\mu \nu} F^{3\,\mu \nu}$ and $\frac{1}{4} F^{8}_{\mu \nu} F^{8\,\mu \nu}$. As a result the Lagrangian takes the following form (we set $\xi=1$ and omit boundary terms): 
\begin{eqnarray*}
& L=\frac{1}{2} \left(D_{\mu} B_{\nu}^p +i \epsilon^{pqr} \overline{B^q_{\mu} B^r_{\nu}} \right)\left( D^{\mu} \overline{B^{p\,\nu}}-i\epsilon^{pqr} B^q_{\mu} B^r_{\nu}\right)- (B^q_\mu \overline{B^q_\mu})^2+ B^r_\mu \overline{B^q_\mu} B^q_\nu \overline{B^r_\nu}+\nonumber\\& +\frac{1}{4} \left(\partial_{[\mu} A^3_{\nu]}+2\Pt_{\mu\nu} \right)\left(\partial^{[\mu} A^{3\,\nu]}+2{\Pt}^{\mu\nu} \right)+\frac{1}{4} \left(\partial_{[\mu} A^8_{\nu]}+2\Pe_{\mu\nu} \right)\left(\partial^{[\mu} A^{8\, \nu]}+2{\Pe}^{\mu\nu} \right)-\frac{3}{4}\Pt_{\mu\nu}{\Pt}^{\mu\nu}-\frac{3}{4}\Pe_{\mu\nu}{\Pe}^{\mu\nu},
\end{eqnarray*}
where
\begin{eqnarray*}
 \Pt_{\mu\nu}=-\iu  \sqrt{2} B^1_{[\mu} \overline{B^1_{\nu]}}+\iu\frac{\sqrt{2}}{2} B^2_{[\mu} \overline{B^2_{\nu]}}+\iu\frac{\sqrt{2}}{2} B^3_{[\mu}\overline{B^3_{\nu]}},\quad \Pe_{\mu\nu}=\iu\frac{\sqrt{6}}{2} B^2_{[\mu} \overline{B^2_{\nu]}}-\iu\frac{\sqrt{6}}{2} B^3_{[\mu}\overline{B^3_{\nu]}}.
\end{eqnarray*}

\end{document}